\documentclass[conference]{styles/IEEEtran}

\usepackage{calrsfs}
\DeclareMathAlphabet{\pazocal}{OMS}{zplm}{m}{n}
\usepackage{epsfig,endnotes,xspace,hyperref,url,diagbox}
\AtBeginDocument{}
\usepackage{amsmath,amsthm,amssymb}
\usepackage{tikz}
\usepackage{enumerate}
\usepackage{varioref}
\usepackage{graphicx}
\usepackage{epstopdf}
\usepackage{color}
\usepackage{xcolor}
\usepackage{colortbl}
\usepackage{multirow}
\usepackage{subfigure}
\usepackage{comment}
\usepackage[utf8]{inputenc}
\usepackage{mathtools}
\usepackage{wrapfig}
\usepackage{algorithm}
\usepackage[noend]{algpseudocode}

\usepackage{caption}
\usepackage{rotating} 
\usepackage{enumitem}

\usepackage{footnote}
\usepackage{ulem}
\makesavenoteenv{tabular}

\definecolor{revision}{RGB}{0,0,0}
\newcommand{\revisiondel}[1]{}
\newcommand{\revision}[1]{\textcolor{black}{#1}}
\renewcommand{\sout}[1]{}
\newcommand{\revisionstart}{\begin{color}{revision}}
\newcommand{\revisionend}{~\!\!\end{color}}

\newcommand{\mypara}[1]{\vspace*{0.05in}\noindent\textbf{#1}$\;$}

\newtheorem{definition}{Definition}

\renewcommand{\Pr}[1]{\ensuremath{\mathsf{Pr}\left[#1\right]}\xspace}

\renewcommand{\AA}{\mathbf{A}}

\renewcommand{\Pr}[1]{\ensuremath{\mathsf{Pr} \left[#1\right] }\xspace}

\newcommand{\methodfull}{Towards Effective Differential Privacy Communication for User\revision{s'} Data Sharing Decision and Comprehension}

\newcommand{\results}{\ensuremath{\mathsf{\mathbf{R}}}\xspace} 
\title{\huge \methodfull}

\begin{document}

\author{

	{\rm Aiping Xiong}\\
	Penn State University
	\and 
	{\rm Tianhao Wang}\\
	Purdue University
	\and
	{\rm Ninghui Li}\\
	Purdue University
	\and
	{\rm Somesh Jha}\\
	University of Wisconsin-Madison
}

\maketitle

\begin{abstract}

	\revisionstart
	Differential privacy protects an individual's privacy by perturbing data on an aggregated level (DP) or individual level (LDP).  We report four online human-subject experiments investigating the effects of using different approaches to communicate differential privacy techniques to laypersons in a health app data collection setting.  Experiments 1 and 2 investigated participants' data disclosure decisions\sout{of} \revision{for} low-sensitive and high-sensitive personal information when given different DP or LDP descriptions.  Experiments 3 and 4 uncovered reasons behind participants' data sharing decisions, and examined participants' subjective and objective comprehensions of these DP or LDP descriptions.  When shown descriptions that explain the implications instead of the definition/processes of DP or LDP technique, participants demonstrated better comprehension\sout{,} and showed more willingness to share information with LDP than with DP, indicating their understanding of LDP's stronger privacy guarantee compared with DP.
	\revisionend

\end{abstract}

\section{Introduction}

The proliferation and ubiquitousness of pervasive computing has brought an unprecedented amount of collection and analysis of personal information.  While such data can be used for personal and societal benefit, improving sustainability, public health, etc\textcolor{cyan}{.}, they can also be used in undesired and unexpected ways.  These usages can cause adverse consequences for data participants' reputation, insurability, etc., leading to hosts of privacy concerns. People distrust current tools~\cite{melicher2016not}, and utilize a variety of measures to protect privacy~\cite{kang2014privacy,sannon2018privacy}, such as withholding personal information or deliberately providing false personal information, which is detrimental to the utility of the collected data.

To protect data privacy and ensure utility in the context of data publishing, the concept of differential privacy (DP) has been proposed~\cite{Dwo06}, which adds noise to the aggregated result such that the amount of revealed information for any individual is bounded.  DP techniques have been deployed by government agencies such as the US Census Bureau for the 2020 census~\cite{uscensus}.
In recent years, local differential privacy (LDP) has been proposed.  LDP differs from DP in that random noise is added at \revision{an} individual user level before sending the data to the server.  Thus, under LDP users do not need to rely on the trustworthiness of the company or the server.   LDP has been deployed by companies such as Google~\cite{rappor}, Apple~\cite{apple-dp}, and Microsoft~\cite{nips:DingKY17}.
With the increasing deployment of DP and LDP techniques, an interesting and important open question is whether users understand these techniques, trust them, and\textcolor{cyan}{,} consequently, increase their data disclosure when these techniques are deployed.

Our work takes a step towards understanding how to \textit{effectively} communicate DP and LDP techniques \revision{in order} to facilitate users' data disclosure decisions. Centering on textual descriptions of differential privacy techniques, we set out to answer the following five research questions (RQs):

\begin{itemize}[leftmargin=*]
	\item \mypara{RQ 1:}
	      Will participants increase their data disclosure, especially for high-sensitive information, when informed that DP or LDP techniques have been deployed?

	\item \mypara{RQ 2:} To what extent will participants' data disclosure decisions depend on how the privacy techniques are communicated, e.g., description\revision{s} focus\sout{ing} on definition or implication?

	\item \mypara{RQ 3:} What factors caused participants to decide whether or not to share their personal information when given description(s) of differential privacy?

	\item \mypara{RQ 4:}\sout{Whether}\revision{Do} participants feel that they understand the description(s)?  \sout{And}\revision{Moreover,} which part\revision{(s)} \sout{or parts }of the description(s) \sout{are}\revision{is} difficult for them to understand?

	      \revisionstart
	\item \mypara{RQ 5:} In which ways are participants' objective comprehension of DP and LDP affected by how the techniques are described?\revisionend

\end{itemize}

\begin{figure}[t]\centering
	\includegraphics[width=0.47\textwidth]{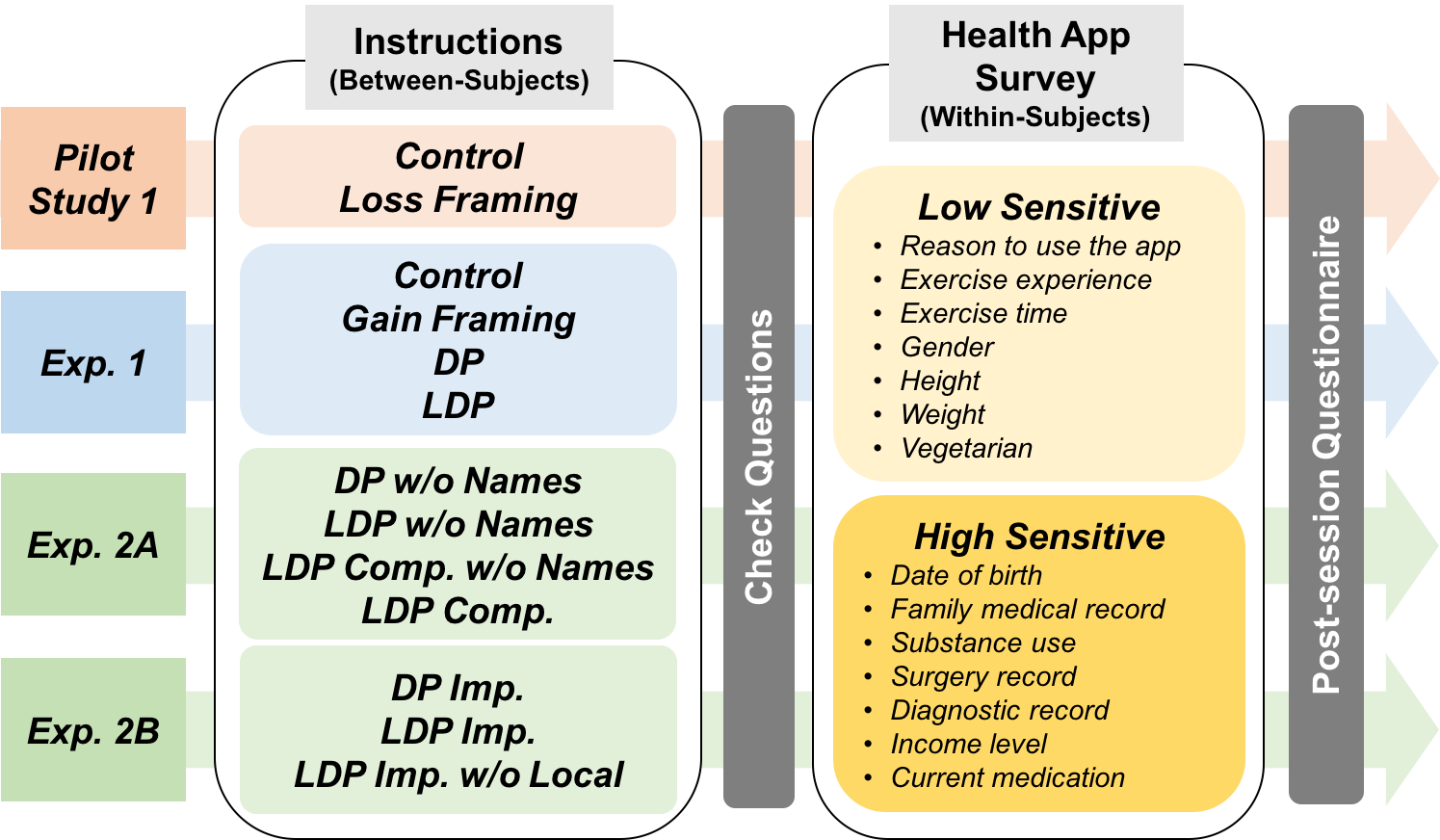}
	\caption{\small{Flow chart shows the experimental design for experiments \revision{of data sharing decisions (Group 1).} \sout{in Group 1.}  Pilot Study 1 validated the health app data collection setting.  Experiments 1, 2A, and 2B addressed RQ1 and RQ2.  ``Instructions'' box presents the conditions in all experiments.  ``Health App Survey'' box lists the seven low-sensitive and the seven high-sensitive questions.  \sout{Experiments 3 and 4 used different procedures. Experiment 3 addressed RQ 3 and RQ4.  Experiment 4 addressed RQ5.} }}
	\label{fig:flow}
	\vspace{-0.3cm}
\end{figure}

\textbf{RQ1} and \textbf{RQ2} are about users' \textit{data sharing decisions} when informed that DP and LDP techniques have been deployed.  To address them, we conducted Experiments 1 and 2, in which participants\sout{were asked to make} \revision{made} hypothetical data disclosure decisions in a health app survey setting (see Fig.~\ref{fig:flow}).  Participants were asked to imagine that they just installed\sout{a} \revision{the} health app, which needs to collect personal information from them.  They were then shown $14$ questions asking for personal information, among which half are considered high\revision{-}sensitive, and the other half are low\revision{-}sensitive.  We varied the presence and absence, as well as the ways of describing privacy techniques.

We ask \textbf{RQ3}, \textbf{RQ4}, and \textbf{RQ5} to understand the \textit{reasons behind users' data sharing decisions}.
To address \textbf{RQ3} and \textbf{RQ4}, we conducted an open-question survey in Experiment 3.  The procedure was similar to prior experiments; however, participants made only one high-sensitive data disclosure decision.  Following the data disclosure decision, we asked each participant to explain \textit{why} they decided to share or not share their personal information.  Participants also rated whether the given differential privacy description was easy to comprehend.  For participants who indicated the description was not easy to understand, we asked them to highlight the part or parts that were difficult to comprehend.
\revisionstart
We conducted Experiment 4 to \sout{\textit{objectively}} assess participants' \revision{\textit{objective}} comprehension of DP and LDP (\textbf{RQ5}). Based on lessons learned from previous experiments, we also added new descriptions for each technique that explain data flow and implication inferences. Participants were shown descriptions of DP or LDP, then\sout{and asked to } answer\revision{ed} five questions about the privacy and utility consequences.  We compared the correct answer rates of those questions between the two new descriptions and descriptions from prior experiments.
\revisionend

DP and LDP are typically used in different settings.  Each participant in our study was exposed to only one of them.  In terms of privacy, LDP provides stronger protection than DP does, because LDP does not need to trust the server.  Our experimental setting (the health app data collection testbed, see Section~\ref{subsec:exp_overview_procedure}) also focused on the privacy protection provided by the techniques to\sout{eliminate} \revision{minimize} other confound factors in the experiment design.  In this case, we expect a higher data disclosure rate under LDP.
When that did not happen, it could be an indicator that many participants do not really understand the nature of protection from the descriptions.
We obtained answers for each question as follows:

\textbf{RQ1:} \textit{Data Sharing under Differential Privacy.} Participants increased\sout{their} data sharing for \revision{the} high-sensitive questions when they were informed of protection from differential privacy, indicating a positive effect of communicating privacy techniques.

\textbf{RQ2:} \textit{Data Sharing with Different Descriptions}. When descriptions focused on definition and/or data perturbation processes, participates' data sharing\sout{ratio was} \revision{rates were} not larger for LDP than for DP.
\sout{But}\revision{Nevertheless,} higher data disclosure rates were obtained for the LDP conditions than for the DP condition when the implications were communicated, i.e., whether the privacy protection relies on the trustworthiness of the company or the server.

\textbf{RQ3: } \textit{Reasons behind Sharing Decisions.}
About half of the participants chose to share their personal information. Most of them explained that they made the decision because of the described privacy protection\revision{, suggesting}\sout{. This suggests} that trust in privacy protection techniques led to \revision{the} decisions to share.
Participants who decided not to share cited various concerns, top three of which are 1) the requested information are too sensitive to share, 2) distrust of the described differential privacy techniques,
and 3) risks of data breach in the future.

\textbf{RQ4:} \textit{Subjective Measure of Comprehension}.
Only $13\%$ \revision{of the}
participants indicated that they had difficulty in understanding the described techniques, and the\sout{most mentioned} difficult parts \revision{mentioned the most} were about \revision{the} data perturbation processes.

\revisionstart

\textbf{RQ5:} \textit{Objective Measures of Comprehension}.
Better comprehension results were obtained for descriptions that provide implication inferences than those which do not.
\revisionend

Finally, we discuss how the obtained results inform our understanding of effective differential privacy communication\sout{,} and highlight implications of our findings in Section~\ref{sec:discussion}.

To summarize, our work makes the following contributions:

\begin{itemize}[leftmargin=*]

	\item We provide \textit{quantitative} and \textit{qualitative} evidence showing benefits (increasing data sharing) of communicating differential privacy to\sout{laypeople} \revision{users}.\vskip 0.02in

	\item We identify the data perturbation processes as the most difficult parts for laypeople to understand\sout{,} and provide evidence\sout{, i.e., larger data disclosure rates and better objective comprehension results, } showing implication descriptions as one effective way \revision{(i.e., larger data disclosure rates and better objective comprehension results)} for DP and LDP communication.
	      \vskip 0.02in

	\item We further uncover the effect of implication descriptions on comprehending differential privacy with data flow descriptions which afford privacy and utility implication inferences.
	      \vskip 0.02in

	\item We reveal a robust effect of information sensitivity in participants' data disclosure decisions even with privacy\revision{-}enhancing techniques, suggesting biased responses within non-mandatory data collecting.

\end{itemize}

\section{Background and Related Work}
\label{sec:back}

\subsection{Background on Differential Privacy}

\revisionstart
Differential Privacy~\cite{DMNS06} applies in the setting where there is a trusted data curator, who gathers data from individual users, processes the data in a way that satisfies DP, and then publishes the results.
\begin{definition}[Differential Privacy] \label{def:dp}
	An algorithm $\AA$ satisfies $\epsilon$-DP, where $\epsilon \geq 0$,
	if and only if for any two datasets $D$ and $D'$ that differ in at most one record, and any set \results of possible outputs of $\AA$, we have
	\begin{equation*}
		\Pr{\AA(D)\in \results} \leq e^{\epsilon}\, \Pr{\AA(D') \in \results}
	\end{equation*}
\end{definition}
The definition prevents a strong adversary who knows all but one record in the database from inferring the last one after seeing the output.
To ensure that, $\AA$ first obtains the true result from $D$, and then adds noise to the result.

In the local setting, each user perturbs the input value $v$ using an algorithm $\AA$ and reports $\AA(v)$ to the aggregator.

\begin{definition}[Local Differential Privacy] \label{def:ldp}
	An algorithm $\AA(\cdot)$ satisfies $\epsilon$-local differential privacy ($\epsilon$-LDP), where $\epsilon \geq 0$,
	if and only if for any input $v, v'$, and any set \results of possible outputs of $\AA$, we have
	\begin{equation*}
		\Pr{\AA(v)\in \results} \leq e^{\epsilon}\, \Pr{\AA(v')\in \results}
	\end{equation*}
\end{definition}
In both DP and LDP, $\epsilon$ plays an important role as it measures the randomness of the process.  A large $\epsilon$ leads to insufficient noise, which does not provide much privacy protection.

\mypara{Difference between DP and LDP.}
In DP, the server has access to the true sensitive values of the users, while in LDP, the aggregator does not see the actual private data of each individual.  Instead, users send randomized information to the aggregator, who infers the data distribution based on that.  However, the better trust model also comes at the cost of utility: with the same privacy guarantee, measured by the parameter $\epsilon$, the utility of LDP is worse than DP by a factor of $\Theta(\sqrt{n})$~\cite{chan2012optimal}, where $n$ is the number of users.

\revisionend
\vskip -0.15in
\mypara{Deployment of DP and LDP.}
Although DP was proposed \sout{for} more than a decade \revision{ago}, the first public deployments of this concept\sout{is} \revision{are} related to LDP, e.g., companies like
Apple~\cite{apple-dp}, Google~\cite{rappor}, and Microsoft~\cite{nips:DingKY17}.
Exemplary use cases include collecting users' default browser homepage and search engine to understand the unwanted or malicious hijacking of user settings; or gathering frequently typed emojis and words to help predict keyboard typing.

More recently, DP is also deployed in both industry and government.  In particular, Uber released an open\revision{-}source project for SQL query with differential privacy~\cite{chorus}; LinkedIn proposed a system to analyze user information with DP~\cite{pripearl};
the US Census Bureau has deployed DP technologies for the 2020 census~\cite{uscensus};
and Harvard built a system prototype for researchers to share data using DP~\cite{psi}.

\revisionstart
With the deployments of DP and LDP, we started seeing companies and organizations begin to communicate DP and LDP techniques to the public, including Apple, Google, Microsoft, Uber, and US Census Bureau. We took descriptions from the\sout{above mentioned} companies and organizations \revision{mentioned above}, made minor modifications to fit our context, and used them in our experiments.  \sout{This helps understand whether these descriptions are effective in communicating DP and LDP to end users.}
\revisionend

\subsection{Related Work}

\revisionstart
\mypara{Usability of DP and LDP.}
One primary goal of addressing the RQs is to achieve ``usable differential privacy''.
The most closely related prior work is Bullek et al.~\cite{bullek2017towards}, which studied people's comprehension of the random response method~\cite{Warner65} for LDP and preference of the privacy parameter.  In that study, each participant perturbed answers for sensitive questions with three probabilities, corresponding to three $\epsilon$ values.  For a final high-sensitive question, participants were asked to first choose the perturbation probability and then answer.  75\% of the participants chose the largest perturbation (which obscured their true answers the most), 5\% chose the intermediate one, and 20\% chose the least perturbation.  Subjective reasons provided for selecting the least protection focused on a desire to respond truthfully.  One interpretation of these results is that most people want strong privacy protection, but a cognitive bias to equate such data perturbation with ``lying'' (data-obfuscation) can sway privacy-related decisions.

Our work is orthogonal to~\cite{bullek2017towards}.  Instead of focusing on the parameter value, we strive to convey the qualitative nature of differential privacy.  We study people's willingness to share personal information when given different descriptions of DP or LDP, and the reasons behind those decision\revision{s}.  While the privacy parameter $\epsilon$ critically affects the level of privacy protection, it seems unlikely that end users can choose $\epsilon$ based on understanding how the mechanism works and assess the impacts of different $\epsilon$ values.  It is more likely that they will rely on expert assessment of the appropriateness of deployed $\epsilon$ values, and accessible explanations of the consequences.  This is similar to how privacy policies are used\sout{in practice}. \revision{In practice, people are not reading privacy policies because they are long and full of legalese~\cite{mcdonald2008cost}. At the same time, when a company's privacy policies and practices are inadequate, this will often be discovered by experts, and lawsuits may ensure~\cite{cadwalladr2018revealed,mills2006aol}. }
\revisionend

\revisionstart
\mypara{Usability of Privacy Notices.}
A large body of work has been conducted to inform users about privacy techniques~\cite{kelley2009nutrition,kelley2010standardizing} and to facilitate their privacy decisions~\cite{gates2014effective,tsai2011effect}.  For example, when privacy policies of online shopping sites were made more prominent and accessible within a shopping search engine interface, participants increased their purchase intention with the sites offering better privacy protection~\cite{tsai2011effect}.

To improve the usability of privacy from the user's perspective, ``privacy by control'' through notice and choice have become essential for privacy protection~\cite{schaub2017designing}.  Notice and consent as a principle is widely recognized by law and \sout{the} society.  For example, companies such as Google and Apple have implemented permission dialog\revision{s} in Android and iOS to request access to hardware and data from users.
Felt and colleagues~\cite{felt2012android} investigated the effectiveness of \revision{the} Android permission system in warning users of app installation risks.  Their results showed that most participants did not pay attention to permission warnings or did not understand what the permissions mean.   Lin et al.~\cite{lin2012expectation} examined permission warnings in helping participants manage \revision{the} privacy of installed apps.  Their results showed that designs that highlight privacy implications, e.g., unexpected data collection practices, were effective in helping participants avoid intrusive apps.

The communication of DP or LDP also deals with the usability of privacy notice.  But it has\sout{the} \revision{a} unique challenge because of its mathematical complexity.  Thus, we start from definitions and explore how to remove the technical details while preserving the fidelity of the communication.
\revisionend

\mypara{Decision Making of Online Data Sharing.}
Our study centers on people's data sharing decisions, which are affected by various factors.
People's decision making in\sout{the} risk contexts are influenced by\sout{the way in which} \revision{how} a problem is framed~\cite{tversky1981framing,tversky1986rational}.
Specifically, if the outcomes are described in term\revision{s} of potential loss (negative framing), people are risk-seeking.  However, people are risk-averse when the outcomes are presented in terms of potential gains (positive framing).

A different way of establishing a frame of reference involves emphasis framing, which accentuates a subset of potential\revision{ly} relevant considerations~\cite{entman1991framing}. For example, the consequence of online data sharing can be framed positively in term\revision{s} of free product and service, or negatively in term\revision{s} of loss due to privacy concern\revision{s}.  While some prior\sout{study} \revision{studies} provided evidence that the emphasis framing influenced people's privacy decisions~\cite{adjerid2013sleights}, some did not~\cite{gluck2016short}, requiring further research.  A scrutiny of the prior studies revealed differences\sout{of} \revision{on} information sensitivity level. While highly intrusive information, such as drug use, was asked in~\cite{adjerid2013sleights}, most information examined in~\cite{gluck2016short}, such as height and time of exercise, were less sensitive.

\revisionstart
Privacy issues arise in the specific contexts~\cite{chen2013privacy,naeini2017privacy,nissenbaum2009privacy}. Prior studies revealed that many\sout{of} health apps had privacy risks to users~\cite{dehling2015exploring,huckvale2015unaddressed}\revision{,} and caused low engagement of users due to privacy concern~\cite{krebs2015health,torous2018clinical}.  So we chose a health app survey setting as \revision{the} testbed to evaluate participants' data sharing decisions in \revision{the} current study.
While DP provides better utility, we note that this is mainly beneficial to the server, rather than the users. \sout{As suggested by the results of~\cite{bullek2017towards}, LDP is the ideal choice for a rational user as it provides better privacy promise than DP does in the same health app data sharing context.} \revision{LDP provides better privacy promise than DP does in the health app data sharing context, which would be preferred by users~\cite{bullek2017towards}.}
We also varied the sensitivity level of the health information across survey questions\sout{,} and evaluated \revision{the} effect of a negative framing in term\revision{s} of privacy risk or a positive framing in term\revision{s} of benefit on participants' data-sharing decisions.
\revisionend

\section{Overview of Experiment Design}
\label{sec:exp_overview}

We ran a series of online experiments, which can be divided into two groups.
Experiments in Group 1 (including Pilot Study 1 and Experiments 1, 2A, 2B) focus on \textit{decision}
measures of whether participants were willing to share their personal information under different conditions.  Experiments in Group 2 (including Experiments 3, Pilot Study 2\revision{,} and Experiment 4) focus on more in-depth understanding \revision{of} the \textit{reasons} behind participants' data sharing decisions and their \textit{comprehension} of DP and LDP.
We ran multiple studies in part because findings in earlier studies led to interesting questions \sout{which} \revision{that} we sought to answer with additional studies.

In this section, we describe the method of participant\sout{s} recruitment, design of differential privacy descriptions, and the testbed of health app data collection.
Experiments in Group 1 all used the same procedure\revision{,} which is described in Section~\ref{sec:exp2}.  Experiments in Group 2 used different task procedures, which are explained in Sections~\ref{sec:exp4} and~\ref{sec:exp5}, respectively.

\subsection{Participant Recruitment}
All experiments were conducted on Amazon Mechanical Turk (MTurk), and the human intelligent task (HIT) was posted with restrictions to US workers with at least 95\% approval rate and 100 or more approved HITs.  We made these restrictions in the studies to accurately represent sample restrictions of most recent MTurk research~\cite{hauser2016attentive}.
All experiments complied with the American Psychological Association Code of Ethics and were approved by the Institutional Review Board at the authors' institutes.  Informed consent was obtained for each participant.  \revision{Data of t}\sout{T}he experiments\sout{data} were anonymized before analysis.

\subsection{Differential Privacy Communication Design}
\label{sec:design_comm}

To come up with the descriptions of DP and LDP to be used in the study, we started from the descriptions published by the companies and organizations that deployed these techniques, and then conducted multiple rounds of internal discussion and review of the descriptions.  In the discussions, we involved experts of differential privacy to ensure that our descriptions of DP and LDP are technically accurate, and laypeople to help ensure that they can be understood.  As mathematical rigour is one key strength of DP and LDP, we decided not to shy away from using mathematical terms such as ``probability" or ``aggregated data" in some of the descriptions. \revision{The full descriptions of all conditions used in the experiments are given in Table~\ref{tbl:tech} from Appendix~\ref{sec:add_result}.
}

To verify and enhance participants' understanding, we added one check question asking participants to recognize the presented technique immediately after each textual description (\revision{see Appendix~\ref{sec:all_ins}}).  For participants who did not answer the question correctly, we presented the corresponding description again.  We asked the same check question in the  post-session questionnaire\sout{to evaluate the effect of the extra presentation.} \revision{evaluating the effect of the second presentation}.

\sout{The full descriptions of all conditions used in the experiments are given in Table~\ref{tbl:tech} from Appendix~\ref{sec:add_result}.}
\subsection{Heath App Data Collection Setting }
\label{subsec:exp_overview_procedure}
For each experiment, we presented the same health app data collection scenario in which\sout{participants were} \revision{each participant was} instructed to play the role \revision{of an health app user} in three steps \revision{(see Appendix~\ref{sec:all_ins} for the details)}.\sout{We} \revision{Within the instructions, we present examples of collected data (e.g., age and gender) at the local app and the app server to let participants better understand the health app data collection and then situate themselves in the hypothetical setting we created.}  \sout{three steps:}

\section{Experiment 1}
\label{sec:exp2}

Before Experiment 1, we conducted Pilot Study 1, which used the health app data collection setting described \revision{above}\sout{in Section~\ref{subsec:exp_overview_procedure} without mentioning privacy techniques}.  We had the following findings from Pilot Study 1.  Participants showed less willingness to answer the high-sensitive questions than the low-sensitive ones.  When the loss framing (explaining privacy threat of data sharing) was presented, participants' data disclosure was reduced regardless of question sensitivity, and the reduction was\sout{more} \revision{larger} for \revision{the} high-sensitive questions.  Thus, we obtained the effect of question sensitivity and the\sout{frame} \revision{framing} effect~\cite{adjerid2013sleights,bilogrevic2016if,gluck2016short}, confirming the health app data collection setting and hypothetical willingness to disclose personal information as \revision{a} testbed to evaluate privacy decisions.
See Appendix~\ref{sec:exp1} for additional details of Pilot Study 1.

Since the benefit of data disclosure in lieu of privacy threat is often emphasized in the wild, we chose to mention the benefit of data sharing (i.e.,\sout{use} the gain framing) in Experiment 1.\sout{To} \revision{We} evaluate\revision{d} the effect of communicating differential privacy techniques on participants' data sharing decisions (\textbf{RQ1})\sout{, there were} \revision{with} four between-subjects conditions: \textit{Control}, \textit{Gain Framing}, \textit{DP}, and \textit{LDP}.  We predicted a main effect of question sensitivity as Pilot Study 1. With an emphasis on possible benefit, we expected that participants in the \textit{Gain Framing} condition would become less concerned about their privacy, and thus would increase data sharing compared to the \textit{Control} condition.  With extra privacy protection in the \textit{DP} and \textit{LDP} conditions, participants would increase their data disclosure further, more for the \textit{LDP} condition with better privacy guarantee.

\subsection{Participants and Stimuli}
We recruited $598$ Amazon MTurk workers. Each participant was paid 1 US dollar for completing the study \revision{(median completion time: about 289 seconds)} The payment rate was the same for all experiments except Experiment 3 (see details in Table~\ref{tbl:exc}).
\revision{The descriptions of \textit{DP} and \textit{LDP} used in the study focused on definitions, and we listed the organizations which have implemented the techniques (see Table~\ref{tbl:tech}). }

\sout{Framing and differential privacy communication were both absent in the \textit{Control} condition.
	We emphasized the benefit of sharing personal information at the beginning of the \textit{Gain Framing} condition, in which participants were shown the following text. }
\revisiondel{
	\begin{itemize}[leftmargin=*]
		\item
		      \textit{\sout{In the current information age, everyone faces one question: Will you share your personal information in return for a product, service or other benefits?}}
	\end{itemize}
}

\sout{The descriptions of DP and LDP used in the study focused on definitions, and we listed the organizations which have implemented the techniques (see Table~\ref{tbl:tech}).   Framing and differential privacy communication were both absent in the \textit{Control} condition.}

\subsection{Procedure}

After accepting the HIT, all participants were directed from MTurk to a survey on Qualtrics, and were assigned to one condition randomly.  \revision{At the beginning of all conditions except \textit{Control}, we emphasized the benefit of sharing personal information}. The study\sout{started} \revision{continued} with a goal description.\revisiondel{:
	\begin{itemize}[leftmargin=*]
		\item
		      \textit{\sout{The purpose of this study is to understand what kind of information you are willing to share with a health app, and how you would like your data to be used}}
	\end{itemize}
}
Following the three-step health-app data collection scenario, \revision{the corresponding differential privacy description was presented in the \textit{DP} or \textit{LDP} condition (see Appendix~\ref{sec:all_ins} for the detailed descriptions). Then,}
participants \revision{in all conditions} answered their data sharing decisions for $14$ questions.
Consistent with \cite{adjerid2013sleights,gluck2016short}, we divide them into seven \textit{low-sensitive} questions (i.e., the reason to use the health app, exercise experience, exercise time, gender, height, weight, vegetarian) and seven \textit{high-sensitive} questions (i.e., date of birth, family medical record, substance use, surgery record, diagnostic record, income level, current medication).

\begin{figure}[ht]
	\centering
	\vskip -0.1in
	\includegraphics[width=0.47\textwidth]{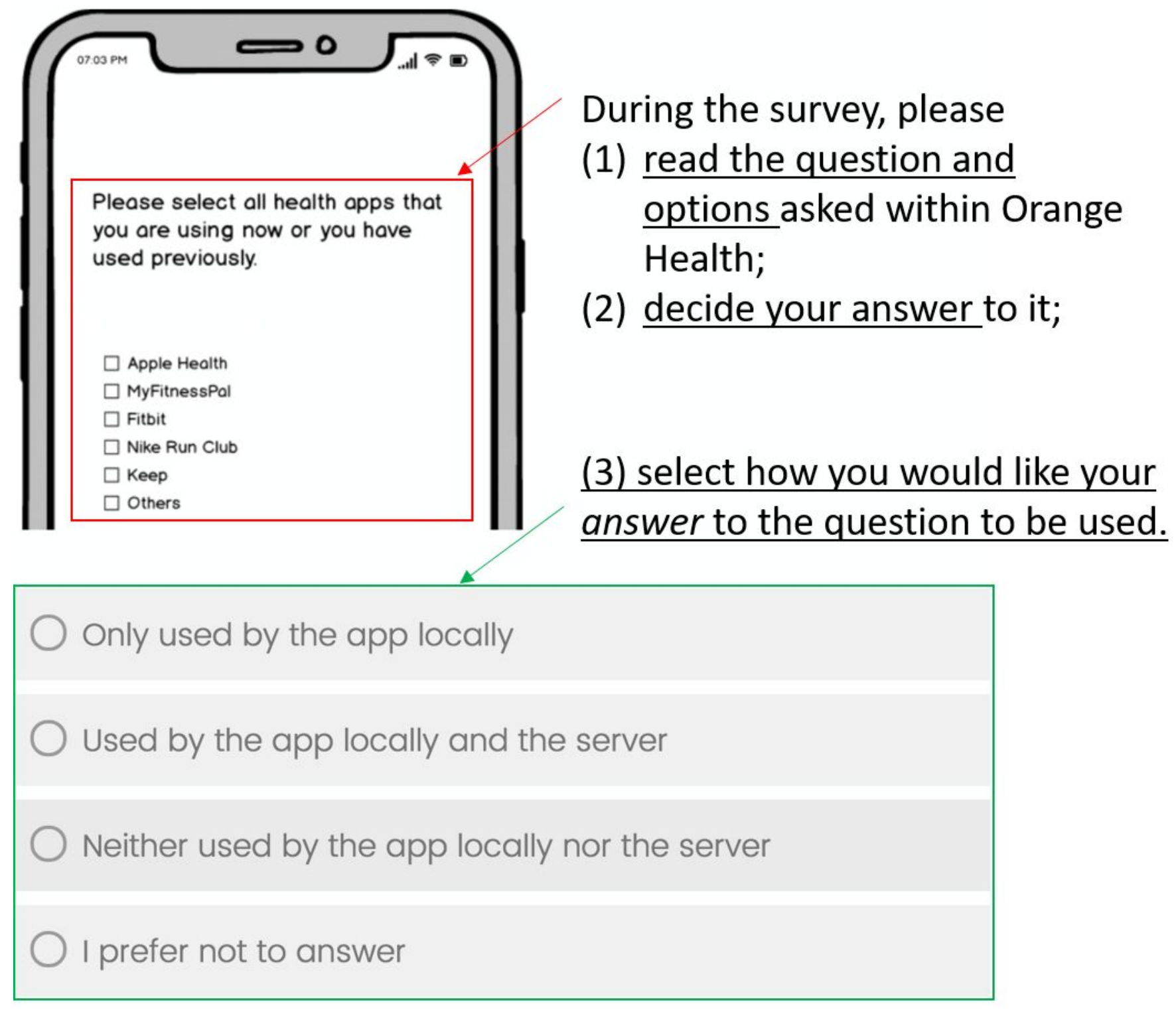}
	\caption{\small{Instructions of survey questions in Experiments 1, 2A and 2B.  ``with DP'' was added to end of options 2 and 3 for all DP related conditions, and ``with LDP'' was added in the same way for all LDP related conditions.}}
	\label{fig:survey_instr}
	\vskip -0.10in
\end{figure}

Each question with its options was presented within a smart-phone layout (see Fig.~\ref{fig:survey_instr}).  Participants were instructed that their task was to read the survey question, decide their answer to it, and select how they would like their answer to be used.  Note that we did not ask participants to actually provide the answers.
We distinguished two types of usage for the collected data, local and at the app server.
So for each question, participants were asked to decide whether they would like the data being used (a) locally only, (b) both locally and at the app server, (c) neither, or (d) prefer not to answer\sout{ (see Appendix~\ref{sec:all_ins})}.

The $14$ questions were presented randomly in each condition.
After answering the $14$\sout{survey} questions, participants completed a questionnaire that asked for demographic information (e.g., age, gender, education, and computer-science background).  We also asked participants to indicate their agreement with statements on their trust for the app and the server on a 7-point Likert Scale (1: strongly disagree, 7: strongly agree), respectively \revision{(see Appendix~\ref{sec:all_ins})}.

For the \revision{\textit{DP}} and \revision{\textit{LDP}} conditions\sout{ with DP or LDP communication},\sout{an extra description of the corresponding technique}
\sout{and the}a check question (see \revision{Appendix~\ref{sec:all_ins}})\sout{were} \revision{was} presented\sout{between the three-step health-app data collection scenario} \revision{after the differential privacy description} and \revision{before} data sharing\sout{decisions} \revision{decision-making.  For participants who did not answer the question correctly, we presented the corresponding description again}.\sout{At the end of the post-session questionnaire, participants answered the check question again and also indicated their trust level for DP or LDP on the 7-point scale.} \revision{Participants answered the check question again at the end of the post-session questionnaire.  They also indicated their trust level for DP or LDP on the 7-point scale.}
We did not obtain any significant difference of the trust evaluations except that there was a main effect of condition in \revision{the} current experiment,
$\chi_{(3)}^2 = 11.44, \textit{p} = .009$.\sout{That p} Participants in the \textit{DP} condition ($65.3\%$) showed more trust than participants in the \textit{Gain Framing} condition did ($50.5\%$), $\textit{$p_{adj}$} = .007$.  Thus, we omit the results of trust evaluation but discuss\sout{it} \revision{them} in the General Discussion (Section~\ref{sec:discussion}).

\subsection{Results}

Participants were excluded from data analysis using two criteria: duplicate IP address and overall completion time less than $120$ seconds.
Due to the main interest in the effect of differential privacy communication, we also excluded participants who did not answer the second check question correctly. \revision{The} \sout{N}number of participants excluded from data analysis were listed in Table~\ref{tbl:exc}.  Consequently, $87$ participants from the \textit{Control}, $101$ from the \textit{Gain Framing}, $150$ from the \textit{DP}, and $127$ from the \textit{LDP}, were included in the data analysis.
The demographic distributions were similar between conditions. See Table~\ref{tbl:demo} for descriptive statistics.

We measured the selected option of each question for each participant.  Decisions were coded as \textit{Opt out} when participants chose ``Neither used by the app locally nor the server'' or ``I prefer not to answer''.
Choices of \textit{Local only} (``Only used by the app locally'') and \textit{Both} (``Used by the app locally and the server''), as well as \textit{Opt out} decisions were determined of each question for each participant.

\begin{figure*}[ht]
	\centering
	\includegraphics[width=1.0\textwidth]{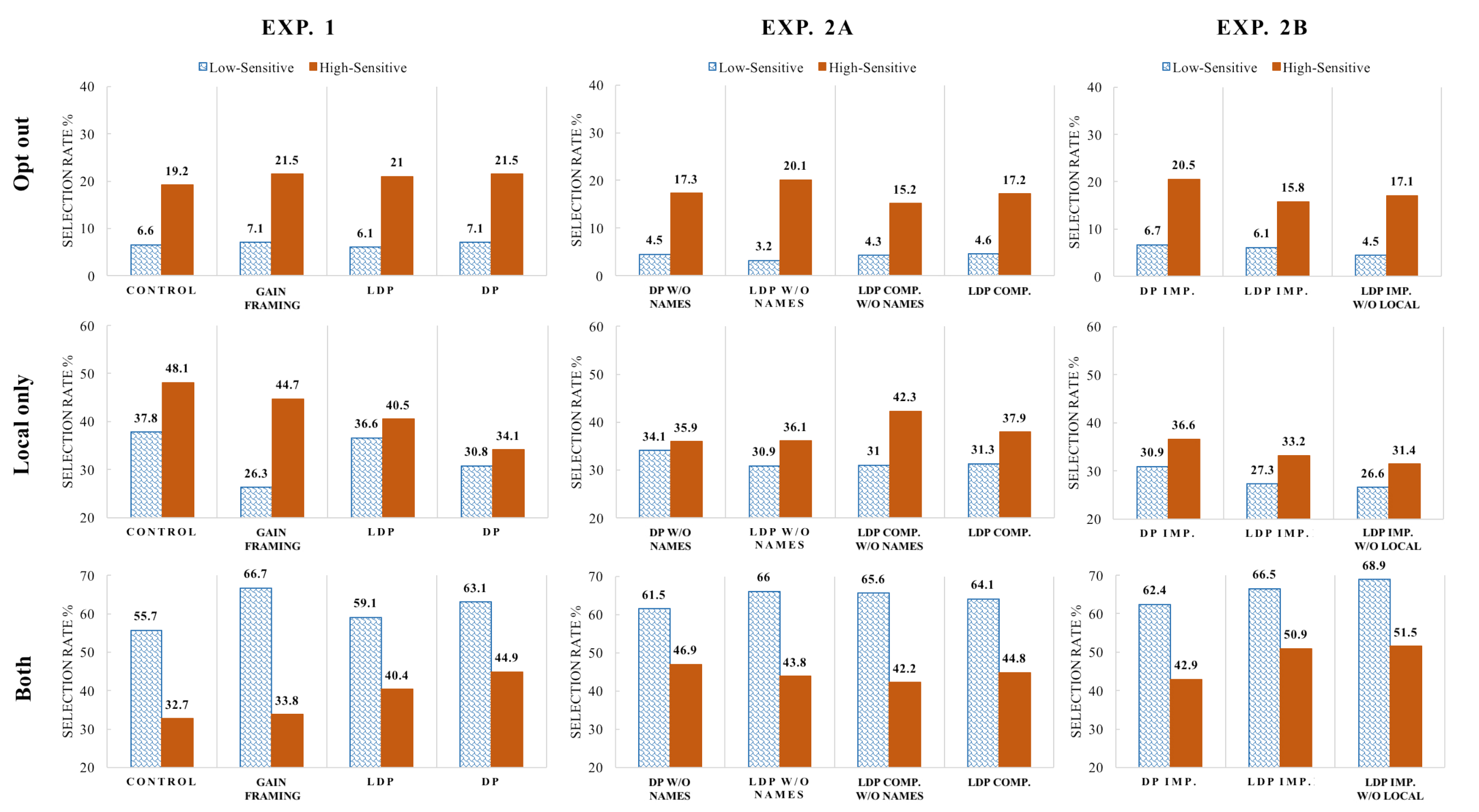}
	\caption{\small{Selection\sout{ratios} \revision{rates} of \textit{Opt-out} decisions (top row), \textit{Local only} (middle row), and \textit{Both} (bottom row) options for \revision{the} low-sensitive and \revision{the} high-sensitive questions in different conditions of Experiments 1, 2A, and 2B. }}
	\vskip -0.15in
	\label{fig:results_revision}
\end{figure*}

\textit{Opt out} decision, selection for \textit{Local only} option, and choice of \textit{Both} option collapsed across participants (see \revision{the} first column of Fig.\sout{ure}~\ref{fig:results_revision}) were entered into a $2$ (question sensitivity: \textit{low-sensitive}, \textit{high-sensitive}) $\times$ 4 (condition: \textit{Control}, \textit{Gain Framing}, \textit{DP}, \textit{LDP}) chi-squared tests with a significance level of .05, respectively.  Post-hoc tests with Bonferroni corrections~\cite{bland1995multiple} were performed, testing all pairwise comparisons with corrected \textit{p}-values for possible inflation.  We report the statistics only for the significant effects in the text.  Please refer to Table~\ref{tbl:stat_1} for the full results of statistical tests.

\begin{table}[ht]
	\begin{center}
		\caption{\small{Demographics of participants in each experiment. \sout{Number in the bracket on the top row indicates the number of participants in each experiment.} \revision{The number of participants of each experiment was listed in the brackets on the top row.} EXP. means Experiment.}}
		\label{tbl:demo}
		\vskip -0.07in
		\resizebox{0.49\textwidth}{!}
		{\begin{tabular}{llccccc}
				\hline
				\multicolumn{1}{c}{\textbf{Item}}                                     & \textbf{Options} & \textbf{\begin{tabular}[c]{@{}c@{}}EXP.1\\ (465)\end{tabular}} & \textbf{\begin{tabular}[c]{@{}c@{}}EXP.2A\\ (581)\end{tabular}} & \textbf{\begin{tabular}[c]{@{}c@{}}EXP.2B\\ (468)\end{tabular}} & \textbf{\begin{tabular}[c]{@{}c@{}}EXP.3\\ (278)\end{tabular}} & \textbf{\begin{tabular}[c]{@{}c@{}}EXP.4\\ (540)\end{tabular}} \\\hline
				\multirow{4}{*}{\rotatebox[origin=c]{90}{Gender}}                     & Male             & 56.8\%                              & 50.4\%                              & 47.9\%                              & 55.0\%                              & 52.6\%                              \\ \cline{2-7}
				                                                                      & Female           & 43.0\%                              & 49.4\%                              & 51.3\%                              & 44.6\%                              & 46.7\%                              \\ \cline{2-7}
				                                                                      & Other            & 0.2\%                               & 0.2\%                               & 0.6\%                               & 0.4\%                               & 0.2\%                               \\ \cline{2-7}
				                                                                      & Not to answer    & 0\%                                 & 0\%                                 & 0.2\%                               & 0\%                                 & 0.6\%                               \\ \hline
				\multirow{6}{*}{\rotatebox[origin=c]{90}{Age}}                        & 18-24            & 8.2\%                               & 7.1\%                               & 6.6\%                               & 10.1\%                              & 8.9\%                               \\ \cline{2-7}
				                                                                      & 25-34            & 47.7\%                              & 42.7\%                              & 44.9\%                              & 35.3\%                              & 34.4\%                              \\ \cline{2-7}
				                                                                      & 35-44            & 24.9\%                              & 28.7\%                              & 29.1\%                              & 26.6\%                              & 23.0\%                              \\ \cline{2-7}
				                                                                      & 45-54            & 10.3\%                              & 13.3\%                              & 11.1\%                              & 15.5\%                              & 16.5\%                              \\  \cline{2-7}
				                                                                      & 55 or older      & 8.6\%                               & 7.9\%                               & 8.3\%                               & 12.6\%                              & 16.9\%                              \\ \cline{2-7}
				                                                                      & Not to answer    & 0.2\%                               & 0.3\%                               & 0\%                                 & 0\%                                 & 0.4\%                               \\ \hline
				\multirow{6}{*}{\rotatebox[origin=c]{90}{Education}}                  & No high school   & 0\%                                 & 0.2\%                               & 0.2\%                               & 0.4\%                               & 0.4\%                               \\ \cline{2-7}
				                                                                      & High School      & 27.7\%                              & 20.5\%                              & 23.7\%                              & 25.9\%                              & 24.6\%                              \\ \cline{2-7}
				                                                                      & College/Bachelor & 60.20\%                             & 65.9\%                              & 62.4\%                              & 59\%                                & 59.1\%                              \\ \cline{2-7}
				                                                                      & Masters/Ph.D.    & 10.8\%                              & 11.9\%                              & 12.0\%                              & 10.8\%                              & 13.9\%                              \\ \cline{2-7}
				                                                                      & Medical degree   & 0.6\%                               & 0.9\%                               & 1.7\%                               & 1.8\%                               & 0.4\%                               \\  \cline{2-7}
				                                                                      & Not to answer    & 0.6\%                               & 0.7\%                               & 0\%                                 & 2.2\%                               & 0.7\%                               \\ \hline
				\multirow{3}{*}{\rotatebox[origin=c]{90}{\begin{tabular}[c]{@{}l@{}}CS \\ Back\end{tabular}}} & Yes              & 22.8\%                              & 19.4\%                              & 28.8\%                              & 20.5\%                              & 20.0\%                              \\ \cline{2-7}
				                                                                      & No               & 75.9\%                              & 79.3\%                              & 70.3\%                              & 76.7\%                              & 78.5\%                              \\ \cline{2-7}
				                                                                      & Not to answer    & 1.3\%                               & 1.2\%                               & 0.9\%                               & 2.9\%                               & 1.5\%                               \\ \hline
			\end{tabular}}
	\end{center}
\end{table}

\subsubsection{RQ1. Effect of Differential Privacy Descriptions}
With extra \textit{DP} and \textit{LDP} descriptions, participants' overall decision rates were similar to that of the \textit{Gain Framing} condition, but they showed more willingness to share high-sensitive information.

\textbf{Opt out rate.} Participants opted out more for the high-sensitive \revision{questions} (20.2\%)
than for the low-sensitive questions (6.0\%), $\chi_{(1)}^2$ $=$ 290, \textit{p} $<$ .001.  Neither the main effect of condition (\textit{Control} vs. \textit{Gain Framing} vs. \textit{LDP} vs. \textit{DP}: 12.9\% vs. 14.3\% vs. 11.8\% vs. 13.5\%), nor its interaction with question sensitivity\sout{were} \revision{was} significant, suggesting little framing effect or the effect of differential privacy communication.

\textbf{Local only selection rate.}
The selection \sout{ratio} \revision{rate} was larger for the high-sensitive questions (41.8\%)
than for the low-sensitive \revision{questions} (32.8\%), $\chi_{(1)}^2 = 43.38, \textit{p} < .001$.  The selection \sout{ratio} \revision{rates} differed across conditions (\textit{Control}: 42.9\%, \textit{Gain Framing} 35.5\%, \textit{LDP} 38.5\%, \textit{DP}: 32.4\%), $\chi_{(3)}^2 = 40.31, \textit{p} < .001$.  Post-hoc analyses revealed \revision{a} significant difference between \textit{Control} and \textit{Gain Framing}, $\textit{$p_{adj}$} < .001$,  suggesting a framing effect.  However, there were no differences\sout{between} \revision{among} \textit{Gain Framing},\sout{ and the two differential privacy description} \revision{\textit{DP}, and \textit{LDP}} conditions.

The two-way interaction of question sensitivity $\times$ condition was\sout{also} significant, $\chi_{(3)}^2 =25.19, \textit{p} < .001$. Post-hoc comparisons showed that participants in the \textit{Gain Framing} and \textit{Control} conditions selected more local option\revision{s} for the high-sensitive \revision{questions} than for the low-sensitive questions, $\textit{$p_{adjs}$} < .001$, but such pattern was not evident in the \textit{DP}\sout{and} \revision{or the} \textit{LDP} condition\sout{s}. Thus, participants preferred high-sensitive personal information to be used by the app locally\sout{generally}, but such preference disappeared when they were informed of \sout{extra description of }DP or LDP.

\textbf{Both selection rate.} Similar\sout{as} \revision{to} the \textit{Local only} option results, main effects of question sensitivity, $\chi_{(1)}^2 = 325.65, \textit{p} < .001$, condition, $\chi_{(3)}^2 = 30.31, \textit{p} < .001$, as well as their interaction, $\chi_{(3)}^2 = 22.45, \textit{p} < .001$, were all significant.  Specifically, \revision{the} selection\sout{ratio} \revision{rate} for the high-sensitive questions (37.9\%) was smaller than that for the low-sensitive questions (61.1\%).  For the average selection\sout{ratio} \revision{rate} of each condition (\textit{Control}: 44.2\%; \textit{Gain Framing}: 50.2\%; \textit{LDP}: 49.7\%; \textit{DP}: 54.0\%),
pairwise comparisons were all significantly different, $\textit{$p_{adjs}$} < .045$, except \revision{for} \textit{Gain Framing} vs. \textit{LDP} and \textit{Gain Framing} vs. \textit{DP}.

Although the effect of question sensitivity was significant for all conditions, $\textit{$p_{adjs}$} < .001$, results of \revision{the} high- and low-sensitive questions showed different patterns across conditions.  For the low-sensitive questions, the selection\sout{ratio} \revision{rate} of the \textit{Gain Framing} condition was larger than that of \textit{Control} and \textit{LDP}, $\textit{$p_{adjs}$} < .013$, but not \textit{DP}. For the high-sensitive questions, the selection\sout{ratio} \revision{rates} for the \textit{DP} and \textit{LDP} conditions were\sout{greater} \revision{higher} than that of \textit{Gain Framing}, $\textit{$p_{adjs}$} <.048$, indicating the effect of differential privacy communication.

\subsubsection{RQ2. DP vs. LDP} Participants showed more willingness to share their information with the \textit{DP} description than for the \textit{LDP} description.

\textbf{Opt out rate.} Results of \textit{LDP} (13.6\%) and \textit{DP} (14.3\%) conditions showed no significant difference.  Similar results were obtained for both \revision{the} low-\sout{sensitive} and high-sensitive questions.

\textbf{Local only selection rate.}
Post-hoc analyses revealed that participants' selection\sout{ratio} \revision{rate} of the \textit{LDP} condition (38.5\%) was larger than that of the \textit{DP} condition (32.4\%), $\textit{$p_{adj}$} < .001$.
Also, such difference was evident for questions of high\sout{-sensitive} \revision{sensitivity}, \textit{$p_{adj}$} $=$ .024, and \revision{of} low\sout{-sensitive} \revision{sensitivity}, \textit{$p_{adj}$} $=$ .047.

\textbf{Both selection rate.} Post-hoc pairwise comparison\revision{s} revealed that the average selection\sout{ratio} \revision{rate} of \textit{LDP} (49.7\%) was smaller than that of \textit{DP} (54.0\%), $\textit{$p_{adj}$} = .045$.  Nevertheless, the selection\sout{ratio} \revision{rate} of the \textit{LDP} condition showed no significant difference from that of the \textit{DP} condition either for the low-sensitive or high-sensitive questions.

\subsubsection{Correct Rate of Check Questions} \revision{The} correct rate of the \textit{DP} condition was higher than that of the \textit{LDP} condition.  Better result\revision{s}\sout{was} \revision{were} evident for the second check question than for the first check question regardless of condition\revision{s}.

Correct answers for check questions collapsed across participants were entered into a $2$ (check: \textit{first}, \textit{second}) $\times$ 2 (condition: \textit{DP}, \textit{LDP}) chi-squared tests.
The\sout{average} correct rate for the \textit{DP} condition (71.6\%) was higher than that of the \textit{LDP} condition (61.8\%), $\chi_{(1)}^2 = 7.60, \textit{p} = .006$, suggesting that the concept of DP was easier to recognize than that of LDP.  The\sout{average} correct rate of the \textit{second} question (74.7\%)
was higher than that of the \textit{first} one (58.7\%), $\chi_{(1)}^2 = 20.53, \textit{p} < .001$, indicating the effect of an extra presentation. The interaction of check $\times$ condition was not significant. Thus the effect of extra presentation played a similar role between the two conditions.

\subsection{Discussion}
Consistent with the results of Pilot Study 1, participants showed more privacy concern\revision{s} for \revision{the} high-sensitive \revision{questions} than for the low-sensitive questions.  The gain framing showed little effect on participants' opt-out decisions, but it increased participants' data disclosure compared to the \textit{Control}.\sout{And} \revision{Moreover,} such increase was only evident for the low-sensitive questions, indicating a risk aversion dependent on the question's sensitivity.

When participants were further informed of DP or LDP protection, their overall data sharing did not increase.\sout{But} \revision{Nevertheless,} they increased their data disclosure for the high-sensitive questions, suggesting a positive effect of communicating DP and LDP. Moreover, the overall data disclosure rate of the \textit{DP} condition was larger than that of the \textit{LDP} condition.  Together with the better results of check questions\sout{ and relatively higher trust}, those results suggest that DP seems to be easier for participants to understand, and thus resulted in more data sharing.

We conjecture that the better results for the \textit{DP} condition may be due to the specific descriptions that we presented. In particular, data perturbation before data collection providing stronger privacy guarantee for LDP was not clearly described.  Also, we included organization names when introducing DP and LDP techniques.  Participants' trust of DP and LDP may depend on their trust of the associated organizations.

\section{Experiments 2A \& 2B}
\label{sec:exp3}
To examine factors that affect DP and LDP description\revision{s} in users' data disclosure decisions, we conducted two sub-experiments. In Experiment 2A, we emphasized the data perturbation process\revision{es} of LDP and examined whether participants would understand the better privacy protection and thus increase their data disclosure.  We also removed the company names associated with DP and LDP to understand their influence on participants' data disclosure decisions.  Considering the difficulty for laypersons to relate data perturbation with privacy protection, we examined the effect of communicating the implications of DP and LDP techniques in Experiment 2B.

\subsection{Experiment 2A}
\subsubsection{Participants, Stimuli, and Procedure}
Another $781$ Amazon MTurk workers were recruited.
To understand the effect of company name, we proposed \textit{DP w/o Names} and \textit{LDP w/o Names} conditions, which were the same as the \textit{DP} and \textit{LDP} conditions of Experiment 1 except company names associated with DP and LDP were removed.  To improve users' comprehension of the better privacy protection provided by LDP, we included a \textit{LDP Comp.} condition, in which the description differed from that of Experiment 1 by emphasizing that data perturbation (noise) was added before sending users' responses to the server.  To further understand the combined influence of \revision{the} company names and the new description, we included a \textit{LDP Comp. w/o Names} condition, which was the same as \textit{LDP Comp.} but the company names were removed.  The detailed descriptions for all conditions are listed in Table~\ref{tbl:tech}.  The procedure was identical to that of Experiment 1.

\subsubsection{Results}
We excluded participants from data analysis using the same criterion as Experiment 1 (see details in Table~\ref{tbl:exc}).
Consequently, $125$ participants from the \textit{DP w/o Names}, $149$ from the \textit{LDP w/o Names}, $161$ from the \textit{LDP Comp. w/o Names}, and $146$ from \textit{LDP Comp.} were included in the data analyses.
Results of each option are shown in the second column of Fig.\sout{ure}~\ref{fig:results_revision}, and were entered into chi-squared tests similar to prior experiment. Post-hoc comparisons were also performed in\sout{the} \revision{a} similar way.

\paragraph{RQ2. Effects of Company Names and Emphasizing Data Perturbation}Without company names, participants' data sharing decisions became similar between DP and LDP descriptions.  In the local setting,  question sensitivity became significant for the \textit{LDP w/o Names} description but not for the \textit{DP w/o Names} description.

\sout{Less} Opt-out rate\sout{for} \revision{of the} high-sensitive question\revision{s} was \revision{less}\sout{evident} for the \textit{LDP Comp. w/o Names} condition than \revision{that} for the \textit{LDP w/o Names} condition.  However, participants in the \textit{LDP Comp. w/o Names} condition preferred to share more high-sensitive information in the local setting than \revision{participants in} the \textit{LDP w/o Names} and \textit{DP w/o Names} conditions.

\textbf{Opt out rate.} Like \revision{the} prior experiment, participants showed more willing to opt out for the high-sensitive \revision{questions} (17.5\%) than for \revision{the} low-sensitive questions (4.1\%), $\chi_{(1)}^2= 372.68, \textit{p} < .001$.  The two-way interaction of question sensitivity $\times$ condition was at .05 significance level, $\chi_{(3)}^2= 7.88, \textit{p} = .048$.  No difference was evident among all conditions for the low-sensitive questions.  For the high-sensitive questions, \revision{the} selection\sout{ratio} \revision{rate} of \revision{the} \textit{LDP w/o Names} condition (20.1\%)\sout{and} was larger \revision{than} that of \revision{the} \textit{LDP Comp. w/o Names} condition (15.3\%), $\textit{$p_{adj}$} = .020$.

\textbf{Local only selection rate.} Participants selected more \textit{Local only} option for the high-sensitive (38.1\%) than for the low-sensitive questions (31.6\%), $\chi_{(1)}^2= 41.54, \textit{p} < .001$.  The interaction of question sensitivity $\times$ condition was significant, $\chi_{(3)}^2= 12.74, \textit{p} = .005$. \sout{That t} The effect of sensitivity was significant for all conditions, $\textit{$p_{adjs}$} < .019$, except for the \textit{DP w/o Names} condition. Thus, when company names were not mentioned, same results as Experiment 1 were evident for DP but not LDP, suggesting the effect of company names in the \textit{LDP} description.  Moreover, post-hoc analysis revealed that selection results across conditions showed no difference for the low-sensitive questions, but for the high-sensitive questions selection ratio of the \textit{LDP Comp. w/o Names} was greater than that of \textit{LDP w/o Names} and \textit{DP w/o Names}, $\textit{$p_{adjs}$} < .017$.

\textbf{Both selection rate.} Similar\sout{as} \revision{to} the other two options, the main effect of question sensitivity, $\chi_{(1)}^2 = 331.56, \textit{p} < .001$, and its interaction with condition, $\chi_{(3)}^2 = 9.06, \textit{p} = .029$, were significant. Participants selected more \textit{Both} option for the low-sensitive \revision{questions} (64.3\%) than for the high-sensitive questions (44.4\%).  Post-hoc analysis showed that the effect of sensitivity was significant for all conditions, $\textit{$p_{adjs}$} < .001$.
Although pairwise comparisons showed no significant differences across conditions for both sensitive levels, different patterns were revealed: For the low-sensitive questions, \revision{the} selection\sout{ratio} \revision{rate} of the \textit{DP w/o Names} condition was numerically smallest.\sout{, whereas,} \revision{In contrast,} for the high-sensitive questions,\sout{the} \revision{its} selection\sout{ratio} \revision{rate} was numerically largest (see Fig.\sout{ure}~\ref{fig:results_revision} second column).

\paragraph{Correct rate of check questions.} Better correct rates were obtained for the two \textit{LDP Comp.} condition\revision{s}.  Same as the prior experiment, participants' correct answer rate for the second check question was better than \revision{that} for the first check question regardless of condition\revision{s}.

Check questions were analyzed in\sout{the} \revision{a} similar way as Experiment 1.
Overall, the correct rates differed across conditions, $\chi_{(3)}^2 = 24.75, \textit{p} < .001$. Post-hoc comparisons revealed that the difference was mainly due to better results of the \textit{LDP Comp. w/o Names} (77.5\%) and \textit{LDP Comp.} (73.9\%) conditions than that of the \textit{DP w/o Names} condition (61.7\%), $\textit{$p_{adjs}$} > .004$, indicating the effect of emphasizing \revision{the} data perturbation process\revision{es}.  Same as Experiment 1, the correct rate of the \textit{second} question (77.8\%) was higher than that of the \textit{first}\sout{one} \revision{question} (63.5\%), $\chi_{(1)}^2 = 36.25, \textit{p} < .001$.

\subsubsection{Discussion}

With an emphasis\sout{e}\sout{of} \revision{on} data perturbation processes, the opt-out rate of \revision{the} high-sensitive questions was smaller for the \textit{LDP Comp. w/o Names} condition than for the \textit{LDP w/o Names} condition.  The check question results of the two \textit{LDP Comp.} conditions were also better than that of the \textit{DP w/o Names} condition.
Instead of increasing their data sharing, participants in the \textit{LDP Comp. w/o Names} condition selected more \textit{Local only} option for the high-sensitive questions than participants in the \textit{LDP w/o Names} and \textit{DP w/o Names} conditions.  Thus, an emphasis\sout{of} \revision{on} data perturbation processes helped participants recognize \revision{the} strong privacy premise of LDP, but it seemed to make them misbelieve that such protection is local and thus showed more willingness to share the high-sensitive information locally.  The effect of ``local'' word was also implied by more selection of \textit{Local only} option in the \textit{LDP} condition than in the \textit{DP} condition in Experiment 1.

After removing \revision{the} company names, the data disclosure differences between \revision{the} \textit{DP} and \textit{LDP} conditions in Experiment 1 were not evident. For the \textit{Local only} option, the non-significant effect of question sensitivity for both \textit{DP} and \textit{LDP} conditions in Experiment 1 became significant for the \textit{LDP w/o Names} condition only. Altogether, those results suggest \revision{the} company names contributed to the differences obtained between \revision{the} \revision{\textit{DP}} and \revision{\textit{LDP} conditions} in Experiment 1, and the company names associated with LDP seemed to have more impact than company names associated with DP.

\subsection{Experiment 2B}
As suggested by the results of Experiment 2A,  participants had difficulty in understanding what do data perturbation processes mean for privacy protection.
Thus, we conducted Experiment 2B examining whether communicating the privacy implication~\cite{lin2012expectation} of data perturbation will help participants understand the stronger privacy promise of LDP.

\subsubsection{Participants, Stimuli, and Procedure}
Extra $600$ Amazon MTurk workers were recruited.
Materials and procedures of Experiments 2B were identical to Experiment 2A except \textit{DP Imp.} and \textit{LDP Imp.} descriptions \revision{(}i.e., whether the privacy protection provided by DP or LDP relies on the trustworthiness of the company or the server\revision{)} were implemented.  We also included\sout{another} \revision{one} \textit{LDP Imp. w/o Local} condition to examine any impact of the word ``local'' on participants' data sharing decisions.  The detailed descriptions are listed in Table~\ref{tbl:tech}.

\subsubsection{Results}
After excluding participants using the same criteria as prior experiments (see details in Table~\ref{tbl:exc}), there were $162$ participants\sout{of} \revision{in} the \textit{DP Imp.} \revision{condition}, $149$\sout{of} \revision{in} the \textit{LDP Imp.} \revision{condition}, and $157$\sout{of} \revision{in} the \textit{LDP Imp. w/o Local} \revision{condition}.  \sout{Each option decisions}\revision{Selections of each option} collapsed across participants (see Fig.\sout{ure}~\ref{fig:results_revision} last column) were analyzed similarly as prior experiments.

\paragraph{RQ2. Effects of Implication Descriptions and Word ``Local''} More data sharing and\sout{less} \revision{fewer} opt out were evident for the two \textit{LDP Imp.} conditions than for the \textit{DP Imp.} condition regardless of questions' sensitivity.  With \revision{the} implication description\revision{s}, no significant difference was evident between \textit{LDP Imp.} and \textit{LDP Imp. w/o Local}.

\textbf{Opt out rate.} Participants opted out more for the high-sensitive \revision{questions} (17.8\%) than for the low-sensitive questions (5.8\%), $\chi_{(1)}^2= 229.49, \textit{p} < .001$.  Opt-out rates also differed across conditions, $\chi_{(2)}^2= 10.77, \textit{p} = .005$, with the obtained result of the \textit{DP Imp.} condition (13.6\%) being larger than those of the \textit{LDP Imp.} (10.9\%) and the \textit{LDP Imp. w/o Local} (10.8\%) conditions, $\textit{$p_{adjs}$} < .026$.  The two-way interaction of sensitivity $\times$ condition was not significant. Thus, across experiments, for the first time\revision{,} we obtained \revision{a} smaller opt-out rate for LDP than for DP regardless of question sensitivity, suggesting the effect of implication communication.

\textbf{Local only selection rate.} Same as the opt-out decisions, only the two main effects were significant. Participants selected more \textit{Local only} option\revision{s} for the high-sensitive \revision{questions} (33.7\%) than for the low-sensitive questions (28.3\%), $\chi_{(1)}^2= 22.58, \textit{p} < .001$.  Selection\sout{ratios} \revision{rates} varied across conditions, $\chi_{(2)}^2= 12.36, \textit{p} = .002$.  Post-hoc pairwise comparisons revealed that the selection \revision{rate} of \textit{DP Imp.} (33.7\%) was larger than those of \textit{LDP Imp.} (30.3\%), $\textit{$p_{adj}$} = .049$, and \textit{LDP Imp. w/o Local} (29\%), $\textit{$p_{adj}$} = .002$\revision{, respectively}.

\textbf{Both selection rate.} Participants decided to share more low-sensitive information (65.9\%) than high-sensitive information (48.4\%), $\chi_{(1)}^2 = 206.02, \textit{p} < .001$.  The main effect of condition was significant, $\chi_{(2)}^2 = 29.2, \textit{p} < .001$.  Post-hoc comparisons revealed that the selection\sout{ratio} \revision{rate} of \textit{DP Imp.} (52.6\%) was\sout{smaller} \revision{less} than\sout{that} \revision{those} of \textit{LDP Imp.} (58.7\%), $\textit{$p_{adj}$} = .016$, and \textit{LDP Imp. w/o Local} (60.2\%), $\textit{$p_{adj}$} < .001$\revision{, respectively}.

\paragraph{Correct rate of check questions} \revision{A larger}\sout{better} correct rate was obtained for the \textit{DP Imp.} condition than for the two \textit{LDP Imp.} conditions.  Same as the prior experiments, participants' correct answer rate for the second check question was better than for the first check question regardless of condition.

Check\revision{-}questions\revision{'} results were analyzed similarly as Experiment 2A.  The correct rates differed across conditions, $\chi_{(2)}^2 = 13.41, \textit{p} = .001$. Post-hoc comparisons revealed that the difference was mainly due to better results of \sout{the} \textit{DP Imp.} (81.6\%) than \revision{those of} \textit{LDP Imp.} (71.6\%)\sout{, and than} \revision{and} \textit{LDP Imp. w/o Local} (71.6\%), $\textit{$p_{adjs}$} < .004$.  But the two \textit{LDP Imp.} conditions did not differ. The correct rate of the \textit{second} check question (81.3\%) was higher than the \textit{first} one (68.6\%), $\chi_{(1)}^2 = 23.95, \textit{p} < .001$. \subsubsection{Discussion}
In Experiment 2B, we made it clear to participants that DP but not LDP relies on the trustworthiness of the company or the server for privacy protection.  When such implications were communicated, more data sharing and\sout{less} \revision{few} opt out were obtained for the two \textit{LDP Imp.} conditions than for the \textit{DP Imp.} condition regardless of question sensitivity.  However,
\revision{the results of check questions}\sout{check question results} revealed that participants did better for answering the DP concept.  Altogether, those results indicate that participants understood\sout{the} better protection provided by LDP, but the concept of DP might still be easier for them to recognize. We did not obtain any difference between the two \textit{LDP Imp.} conditions, indicating \revision{little}\sout{minimal} impact of \revision{the} word ``local'' with implication communication.

\subsection{Summary of Experiments 1 and 2}
Using the health app data collection setting and hypothetical willingness to disclose sensitive personal information as \revision{the}\sout{a} testbed, we evaluated participants' data disclosure rates as a function of differential privacy description.  When definitions of DP and LDP were communicated (Experiment 1), participants increased their data disclosure for\sout{the} high-sensitive information, suggesting a positive effect of communicating differential privacy to laypeople.  However, the overall data sharing was better for DP than for LDP, though the latter provides better privacy guarantee.

When we emphasized the data perturbation processes of LDP (Experiment 2A), participants showed more willingness to share high-sensitive information with the app locally.  However, when the implications of DP\sout{and} \revision{or} LDP (i.e., whether the privacy protection relies on the trustworthiness of the company)\sout{were} \revision{was} presented (Experiment 2B), participants\sout{reduced their opt-out rates} \revision{showed the willingness to opt out less} and\sout{increased their data disclosure} \revision{to share more}\sout{for} \revision{with} LDP\sout{relative to} \revision{than with} DP.  Altogether, those results suggest laypeople have difficulty in understanding the definitions, especially the perturbation processes of differential privacy.  But\sout{a} communication of implication\revision{s} is effective, especially in helping them\sout{to} understand which technique provides better privacy protection.
\section{Experiment 3}
\label{sec:exp4}

\sout{To capture how the differential privacy communication factored into participants' data disclosure decisions, w} We conducted an open-question survey to understand why participants decided to share or not share their personal information given the differential privacy protection (\textbf{RQ3}), and how easy it is for them to understand the descriptions subjectively (\textbf{RQ4}).

\subsection{Participants, Stimuli, and Procedure}
\label{sec:exp4_procedure}
We recruited extra 280 Amazon MTurk workers.
Besides the differential privacy descriptions from prior experiments, we investigated the descriptions published by companies and organizations that implemented DP or LDP, including Apple, Google, Microsoft, Uber, and US Census Bureau.  We made minor changes to those descriptions to make them fit into our study (see descriptions in Table~\ref{tbl:tech} from Appendix~\ref{sec:add_result}).

At the beginning of the survey, participants were instructed that the study is \revision{(1)} to evaluate their data disclosure decision given one privacy protection technique, and \revision{(2)} to understand why they decide to do so.  Then, we described the three steps of role play as prior experiments\sout{(see details in Appendix~\ref{sec:all_ins})}.  Participants were randomly assigned to one of the descriptions.  After viewing the description, participants made one data disclosure decision for high-sensitive information only \revision{(see details in Appendix~\ref{sec:all_ins})}.  Then they answered an open question about the reason for their choice.
We also asked them to indicate their agreement on whether the description was easy to comprehend on a 7-point Likert scale (1: strongly disagree, 7: strongly agree).  For participants who gave ratings less than 4, we asked them to highlight the words or sentences that they thought were difficult to understand.  In the end, participants answered \revision{the same demographic} questions\sout{about their demographics same} as prior experiments.

\subsection{Results}
Only duplicate IP address was used for data exclusion due to short survey time (see Table~\ref{tbl:exc}).  Another participant \revision{who} failed to complete the study was also excluded.
Table~\ref{tbl:demo} lists the summary of participants' demographics.

\subsubsection{Data Disclosure Decision and Difficult-to-Comprehend Rates}
Given descriptions of differential privacy protection, $47.8\%$ of the participants chose to share their high-sensitive information on average (see Table~\ref{tbl:result_exp3}). Across all conditions, the numerically largest sharing\sout{ratio} \revision{rate} was obtained in the \textit{LDP Imp.} condition ($65.2\%$), in agreement with the results obtained in Experiment 2B.   On average, $13.3\%$ of the participants gave a rating less than $4$, indicating that they had difficulty in understanding the presented descriptions.  For conditions included terms, such as ``noise'' or ``random responses'' (e.g., \textit{LDP}, \textit{Google}, \textit{US Census Bureau}), about 18\% of the participants rated the descriptions as difficult (see Table~\ref{tbl:result_exp3})\sout{, whereas}\revision{. In contrast,} for conditions without those terms and mentioned benefit\revision{s} or implications of the techniques (e.g., \textit{DP Imp.}, \textit{Uber}), around $5\%$ of the participants gave ratings less than $4$.  No participants in \revision{the} \textit{DP Imp.} condition\sout{gave ratings less than $4$} \revision{rated the description as hard to understand}, in agreement with better check questions results obtained in Experiment 2B.

\begin{table}[ht]
	\caption{\small{\sout{Ratios} Difficult-to-comprehension and sharing decision \revision{rates} for all descriptions in Experiment 3. \sout{Number in the bracket indicates the number of participants in each condition.} \revision{The number of participants in each condition is listed in the bracket\revision{s} of the first column.}}}
	\label{tbl:result_exp3}
	\begin{tabular}{lcc}
		\hline
		\textbf{Condition}       & \begin{tabular}[c]{@{}c@{}}\textbf{Difficult-to-}\\\textbf{Comprehend Rate}\end{tabular} & \begin{tabular}[c]{@{}c@{}}\textbf{Sharing} \\ \textbf{Decision Rate}\end{tabular} \\ \hline
		DP Imp. (22)             & 0.0\%                      & 36.4\%                     \\ \hline
		LDP Comp. (22)           & 4.5\%                      & 40.9\%                     \\ \hline
		Microsoft (19)           & 5.3\%                      & 52.6\%                     \\ \hline
		LDP Imp. w/o Local (18)  & 5.6\%                      & 44.4\%                     \\ \hline
		Uber (18)                & 5.6\%                      & 55.6\%                     \\ \hline
		LDP Imp. (23)            & 8.7\%                      & 65.2\%                     \\ \hline
		LDP Comp. w/o Names (18) & 11.1\%                     & 50.0\%                     \\ \hline
		Apple (21)               & 14.3\%                     & 57.1\%                     \\ \hline
		LDP w/o Names (19)       & 15.8\%                     & 52.6\%                     \\ \hline
		DP (21)                  & 19.0\%                     & 33.3\%                     \\ \hline
		Google (19)              & 21.1\%                     & 42.1\%                     \\ \hline
		LDP (17)                 & 23.5\%                     & 41.2\%                     \\ \hline
		US Census Bureau (21)    & 23.8\%                     & 47.6\%                     \\ \hline
		DP w/o Names (20)        & 30.0\%                     & 50.0\%                     \\ \hline
	\end{tabular}
\end{table}

\subsubsection{Answers to Open Questions}
We analyzed the results of open questions by identifying themes and generating codes using an inductive approach~\cite{braun2006using}.  The first two authors independently coded the answers for open questions based on the data disclosure decisions and easy-to-comprehend measures, and then cleaned up the codes to generate new ones.  We then re-coded the results using the new codes\sout{,} and added emerging codes when necessary.  Lastly, the research team discussed the codes and grouped them into different themes.  We assigned random sequential numbers to participants for the analysis.

\paragraph{Why share?}
\sout{There were $133$} \revision{One hundred and thirty-three} participants choosing to share personal information.  Their explanations were grouped into three main themes:

\begin{itemize}[leftmargin=*]
	\item
	      \textbf{Trust of DP and LDP techniques.}
	      About 61\% ($82$) of the participants decided to share their information because \revision{of} or partially because of the described privacy protection.  Their replies indicate trust for privacy protection technique\revision{s} generally, such as ``\textit{it sounds like a viable and trustworthy type of protection technique, and I don't feel wary about trusting it.}'' (P124).  Moreover, $28$ of them demonstrated somewhat understanding of differential privacy in their replies, e.g., ``\textit{I think an additional random data set might throw off how certain information can tie to you}'' (P46).

	\item
	      \textbf{Utility consideration.}
	      About 26\% ($34$) of the participants\sout{explained that their decision was} \revision{made the decisions} due to or partly due to their data would be useful or beneficial for the app, the service they got, or\sout{for} the other\sout{s} \revision{people}.  Examples include ``\textit{I feel that I would be able to get more accurate information if it collects my data...}'' (P48),  and ``\textit{If it helps to provide data to make a more accurate algorithm or helps with someone's research I'm willing to provide it}'' (P117). Also, $14$ participants' answers revealed their considerations for both utility and privacy, e.g., ``\textit{I am comfortable to share this information for the benefit that will be served to me. The privacy technique sounds like it will keep all users equally obfuscated}'' (P119).

	\item
	      \textbf{Little privacy concern for asked or any information, learned helpless, and no fear of loss.}
	      About 22\% ($29$) participants explained that they made the decision\revision{s} because (1) they did not care about the privacy for the asked \sout{high-sensitive} information or any information, e.g., ``\textit{...personally I don't currently have a significant history of medical problems, substance abuse, family history, etc...}'' (P113); (2) a lot of their personal information was already out, e.g., ``\textit{we share a lot of info already on social media}'' (P10); or (3) there was nothing to hide or protect, e.g., ``\textit{I don't think its really that big a deal, i tell everyone my business lol}'' (P39).

\end{itemize}

\paragraph{Why not share?}
We also performed a similar analysis to understand why $142$ participants chose not to share, \revision{the results of} which we\revision{re} grouped into \revision{the} following four themes:

\begin{itemize}[leftmargin=*]
	\item
	      \textbf{Too sensitive to share.}  About 37\% ($53$) of the participants\sout{made the not sharing decision} \revision{decided not to share} because of the sensitivity level of requested information, e.g., ``\textit{Because it's personal I have a long list of medical conditions, and my family wouldn't want me to share their personal information as well}'' (P144), and ``\textit{Even if the privacy policy is equivalent to an opt out which might be fine for other circumstances but when you are talking about your health records it just wouldn't be worth the risk.}'' (P224).

	\item
	      \textbf{Distrust differential privacy techniques.} About 33\% ($47$) of the participants explained that they distrust the described differential privacy technique\revision{s}. Besides \revision{the} general concern of privacy techniques, e.g., ``\textit{This technique does not sound like it is full proof. I think a hacker could get through this system real easy.}'' (P154), participants distrust\revision{ed} differential privacy techniques because (1) the descriptions were vague, e.g., ``\textit{I did not understand the explanation of how the protection worked}'' (P200); (2) the techniques were new, e.g.,  ``\textit{Its not tested enough too new.}'' (P167); and (3) further verification is needed, e.g., ``\textit{I'm still not fully convinced that it is trustworthy. I'd need to know more about the processes involved and how thoroughly tested it has been}'' (P191).

	\item
	      \textbf{Risks of data leak, breach, or hack.} About 30\% ($43$) of the participants worried about future risk of leak, breach, or hack of their data, and thus chose not to share, e.g., ``\textit{...All of the data breaches that have occurred in recent years already prove that no privacy protection technique is 100\%. I would like to limit the chances of my info being leaked as much as possible.}'' (P271).

	\item
	      \textbf{Distrust the app or tech companies.} Another 19\% ($27$) participants explained that they chose not to share because they distrust the app or the tech companies, e.g., ``\textit{How good the privacy protection technique is one thing, and whether they will sell my private data to other parties is another thing. I never fully trust such kind of app/technique.}'' (P227).

\end{itemize}

\paragraph{Which part(s) is hard to understand?}
Thirty-seven participants indicated that the descriptions were hard to understand.  They highlighted mostly the words ``\textit{noise}'' ($19$ times) and ``\textit{random(ly)}''($14$ times), both of which are related to the perturbation processes.  When answering why they thought the description was\sout{difficult} \revision{unintelligible}, about half ($17$) of the participants' replies mentioned random noise, including ``\textit{How will the random noise protect my information?...}'' (P90) and ``\textit{what is random noise?}'' (P162). Nine participants also indicated that the descriptions were jargony or had technical terms.

\subsection{Discussion}
Experiment 3 employed an open-question survey to understand factors impacting people's data sharing decisions for high-sensitive information when given descriptions of DP or LDP.
We found that participants decided to share primarily because of
the descriptions of differential privacy techniques and somewhat utility consideration.  Participants who chose not to share had\sout{a variety} \revision{various} concerns: about $1/3$ \revision{of them} believed that the requested information was too sensitive to share, another $1/3$ distrusted the described differential privacy techniques, and an extra $1/3$ worried about negative consequences of sharing high-sensitive information in the future.

We also evaluated participants' subjective comprehension of descriptions to\sout{discover most} \revision{uncover the} difficult parts to understand.
Less than 15\% of the participants rated those descriptions as\sout{difficult} \revision{hard} to comprehend. They highlighted parts closely related to data perturbation processes as most difficult to understand, consistent with results obtained in Experiments 2A and 2B.

\revisionstart
\section{Experiment 4}
\label{sec:exp5}
While participants'\sout{stated} \revision{self-reported} comprehension\sout{of the descriptions} was good \revision{for the descriptions}, it is unclear whether the same holds for their comprehension performance.  To understand participants' \textit{objective} comprehension of DP and LDP (\textbf{RQ5}),
we proposed questions evaluating their understanding of privacy and utility impacts, such as utility cost and who can see their data.

Based on the findings from prior experiments, we also proposed two new descriptions, \textit{DP Flow} and \textit{LDP Flow}.  In each new description, we described the data flow affording privacy and utility implication inferences while simplifying the technical details such as noise and perturbation.  We examined the effect of two new descriptions in improving people's understanding of DP and LDP by comparing \revision{them with}\sout{to} descriptions from prior experiments.

To make sure the two new descriptions and objective comprehension questions are understandable, we conducted Pilot Study 2 with $20$ participants.  The procedure was identical to Experiment 4 except that after answering each question, participants also indicated their agreement on whether the presented question was easy to comprehend on\sout{a} \revision{the} 7-point Likert Scale.  For any question with \revision{a} rating lower than 4, we asked them to describe which part or parts of the question are hard to understand and briefly explain why.   Participants gave overall high ratings for both descriptions and all questions except Q4 \revision{(see details in Appendix~\ref{sec:exp4a})}.  Based on participants' replies, we improved the question
and both descriptions\sout{(see details in Appendix~\ref{sec:exp4a})}.

\subsection{Participants, Stimuli, and Procedure}
We recruited another $599$ Amazon MTurk workers. Descriptions of \textit{DP Flow} and \textit{LDP Flow} are given in Table~\ref{tbl:tech}.  For each description, we introduced DP or LDP data flow by listing different parties, such as external third-party companies, and explaining how those parties are involved \sout{with}in the data flow.  We also added the description\sout{about the loss of accuracy} \revision{of accuracy loss}, and remove\revision{d} technical or jargon terms included in the\sout{prior} \revision{former} proposed descriptions.  Note, in the two new descriptions, \revision{the} risk of data compromise for DP and LDP was not described explicitly. \sout{To investigate the effect of new descriptions,} Descriptions of \textit{DP Imp.}, \textit{LDP Imp.}, \textit{DP w/o Names}, and \textit{LDP w/o Names} were also included \revision{for comparisons with the new descriptions}.

We evaluated participants' comprehension of DP and LDP with five questions.  Three of them were about privacy inferences from \revision{the} perspectives of attackers (Q1), internal employees (Q2), and third-party companies (Q3).  Another two were about utility inferences from the perspectives of the app company (Q4) and third-party companies (Q5) (See Appendix~\ref{sec:exp4a}).

Participants were randomly assigned to one of the six descriptions.  At the beginning, we informed participants that the study was to evaluate their comprehension of one privacy protection technique based on the given description \revision{(see Appendix~\ref{sec:all_ins})}.  Then, we described the three steps of role play\sout{similar} as prior experiments\sout{(see Appendix~\ref{sec:all_ins})}.  After viewing the description of DP or LDP,  participants answered the five questions,
which were presented in a randomized order.  We also randomized the options except ``Unsure'' and ``Prefer not to answer'' for each question.  When answering each question, participants could see the description by placing \revision{their} cursor over the text of ``Hover here to see the description''. \sout{After}Participants \revision{also} indicated whether the description of the privacy protection technique was easy to comprehend on the 7-point Likert Scale.  They answered questions about their demographics in the end.

\subsection{Results}
\revision{The} number of participants in each condition is listed in the\sout{second} \revision{first} row of Table~\ref{tbl:result_exp4}.
Participants' demographic showed a similar pattern as prior experiments (see Table~\ref{tbl:demo}). Correct answer rate of each question for each description collapsed across participants (see Table~\ref{tbl:result_exp4}) were entered into $3$ (description: \textit{w/o Names}, \textit{Imp.}, \textit{Flow}) $\times$ $2$ (technique: \textit{DP}, \textit{LDP}) chi-squared tests.  Post-hoc comparisons were also performed as prior experiments.  Participants' average easy-to-comprehend rating for each description were analyzed with analysis of variance (ANOVA) using the same two factors as chi-squared tests.  Post-hoc tests were also performed with corrected \textit{p}-values for possible inflation.

\begin{table}[ht]
	\caption{\small{Correct answer rate for each question and average comprehension rating of each description in Experiment 4. \sout{Number in the bracket on the second row indicates the number of participants in each condition. Numbers in the bracket on the last row indicates standard error of the average rating.} \revision{The number of participants in each condition is listed in the brackets on the top row. The number in the brackets on the last row indicates the standard error of each average rating.}}}
	\label{tbl:result_exp4}
	\resizebox{0.5\textwidth}{!}{
		\begin{tabular}{lcccccc}
			\hline
			\multirow{2}{*}{\textbf{Question}} & \multicolumn{2}{c}{\textbf{w/o Names}} & \multicolumn{2}{c}{\textbf{Imp.}} & \multicolumn{2}{c}{\textbf{Flow}}                                                                                        \\ \cline{2-7}
			                                   & \begin{tabular}[c]{@{}l@{}}\textbf{DP} \\(\textbf{88})\end{tabular}             & \begin{tabular}[c]{@{}l@{}}\textbf{LDP} \\(\textbf{90})\end{tabular}        & \begin{tabular}[c]{@{}l@{}}\textbf{DP} \\(\textbf{86})\end{tabular}        & \begin{tabular}[c]{@{}l@{}}\textbf{LDP} \\(\textbf{95})\end{tabular} & \begin{tabular}[c]{@{}l@{}}\textbf{DP} \\(\textbf{86})\end{tabular} & \begin{tabular}[c]{@{}l@{}}\textbf{LDP} \\(\textbf{95})\end{tabular} \\ \hline
			Privacy\_Attackers                 & 19.3\%                                 & 25.6\%                            & 87.2\%                            & 76.8\%                     & 51.1\%                     & 77.9\%                     \\ \hline
			Privacy\_Employees                 & 28.4\%                                 & 26.7\%                            & 40.7\%                            & 40.0\%                     & 47.7\%                     & 75.8\%                     \\ \hline
			Privacy\_Third Party               & 52.3\%                                 & 32.2\%                            & 52.3\%                            & 50.5\%                     & 59.3\%                     & 75.8\%                     \\ \hline
			Utility\_Cost                      & 27.3\%                                 & 22.2\%                            & 14.0\%                            & 20.0\%                     & 48.8\%                     & 54.7\%                     \\ \hline
			Utility\_Third Party               & 55.7\%                                 & 60.0\%                            & 81.4\%                            & 56.8\%                     & 89.5\%                     & 84.2\%                     \\ \hline
			\begin{tabular}[c]{@{}l@{}}Easy-to-Comprehend\\ Rating\end{tabular}         & \begin{tabular}[c]{@{}c@{}}4.52 \\(1.71)\end{tabular}             & \begin{tabular}[c]{@{}c@{}}3.53 \\(1.69)\end{tabular}        & \begin{tabular}[c]{@{}c@{}}4.99 \\(1.37)\end{tabular}        & \begin{tabular}[c]{@{}c@{}}4.57 \\(1.53)\end{tabular} & \begin{tabular}[c]{@{}c@{}}4.52 \\(1.59)\end{tabular} & \begin{tabular}[c]{@{}c@{}}5.29 \\(1.25)\end{tabular} \\ \hline
		\end{tabular}
	}
\end{table}

\textbf{RQ5: Effect of Description.} The main effect of description was significant for all comprehension questions, $\chi_{s}^2 > 24.77, \textit{$p_s$} < .001$, (see Table~\ref{tbl:stat_4} for the statistical details).  Across five questions, the best results were obtained for the \textit{Flow} descriptions except Q1, privacy inference of attackers, in which the highest correct\sout{answer} rates were evident for \revision{the} \textit{Imp.} descriptions.  Compared to \revision{the} \textit{Imp.} descriptions, risk of data compromise is implicit in \revision{the} \textit{Flow} descriptions, suggesting the effect of explicitness in helping comprehension.

Correct answer rates of all\sout{private} \revision{privacy-}related questions for the \textit{Imp.} descriptions were larger than those for the \textit{w/o Names} descriptions. Nevertheless, the correct\sout{answer} rates of utility\revision{-}related questions showed no significant difference between those two types of descriptions.  Thus, an implication description\sout{on} \revision{of} data breach is helpful for participants to understand \revision{the} privacy protection of differential privacy.

\textbf{RQ5: Effects of Technique and Technique $\times$ Description.} The main effect of technique showed different patterns among the comprehension questions.  For privacy\revision{-}related questions,  the overall correct rates were\sout{better} \revision{higher} for the LDP conditions than\sout{for} for the DP conditions, $\chi_{s}^2 > 4.09, \textit{$p_s$} < .043$, except \revision{for} Q3, privacy inference about third\revision{-}party companies, which showed no significant difference.  Moreover, the two-way interaction of description $\times$ technique was significant for all three privacy\revision{-}related questions, $\chi_{s}^2 > 10.77, \textit{$p_s$} < .005$.  Generally, \revision{the} difference between DP and LDP was evident for \revision{the} \textit{w/o Names} and \textit{Flow} descriptions, but not the \textit{Imp.} descriptions.

For utility\revision{-}related questions, better \revision{results}\sout{correct answer rates} were evident for DP than for LDP on Q5, utility inference of third\revision{-}party companies, $\chi_{(1)}^2 = 4.07, \textit{p} = .044$.  Also, such pattern was only evident with the \textit{Imp.} descriptions, $\textit{$p_{adj}$} = .002$, indicating the importance of utility description for LDP.

For \textbf{average easy-to-comprehend ratings}, the main effect of description was significant, $F_{(2, 534)} = 27.73, p < .001, \textit{$\eta_{p}^{2}$} = .075$.  Post-hoc pairwise comparisons revealed that the average ratings\sout{for} \revision{of the} \textit{Flow} descriptions (4.90) were similar to those of \revision{the} \textit{Imp.} descriptions (4.78), both of which were higher than those of \revision{the} \textit{w/o Names} (3.94), $\textit{$p_{adjs}$} < .001$.  Although the main effect of\sout{the} technique was not significant (DP vs. LDP: 4.62 vs. 4.47), its interaction with description were significant, \revision{$F_{(2, 534)} = 13.14, p < .001, \textit{$\eta_{p}^{2}$} = .047$}\sout{$F_{(2, 534)} = 13.14, p < .268, \textit{$\eta_{p}^{2}$} = .005$}.  Critically, with \revision{the} \textit{Flow} descriptions, participants' overall rating for LDP\sout{($5.29$)}\sout{became} \revision{was} higher than that for DP ($5.29$ vs. $4.52$), whereas an opposite pattern was evident for both \revision{the} \textit{Imp.} \revision{($4.57$ vs. $4.99$)}
and \revision{the} \textit{w/o Names} \revision{($3.53$ vs. $4.35$)}
descriptions.

\subsection{Discussion}
Contrary to the self-reported results in Experiment 3, objective comprehension results revealed that participants had difficulty in understanding the implications of DP and LDP, especially with descriptions focusing on definition (i.e., \revision{the} \textit{w/o Names} descriptions).  Explicit descriptions about the trustworthiness of the company (i.e., \revision{the} \textit{Imp.} descriptions) improved the correct answer rates for privacy inference questions, especially the inference about attackers.
Participants’ correct answer rates for all questions were improved with the \textit{Flow} descriptions except the inference about attackers.  That is probably because the privacy inference of attackers became somewhat implicit in the \textit{Flow} descriptions compared to the \textit{Imp.} descriptions.  Altogether, those results indicate the effects of explicitness and descriptions affording implication inferences in helping laypeople understand differential privacy.

\subsection{Summary of Experiments 3 and 4}

\revision{Using a similar setting as prior experiments, we asked participants to explain why they decided to share or not share their personal information given the descriptions of differential privacy.  Participants chose to share data mainly because of the described privacy protection, but those who chose not to disclose their personal information revealed different concerns, including the requested information was too sensitive to share, they distrusted the described privacy technique, and they worried about the risk of data breach. Less than 15\% of participants rated the descriptions as hard to comprehend, and they mainly highlighted the parts related to data perturbation processes as difficult to understand.}
\revision{Experiment 4 was conducted to understand how participants comprehend DP and LDP objectively.  Based on the findings from prior experiments, we proposed the \textit{Flow} descriptions which afford privacy and utility implication inferences.  Compared to the \textit{Imp.} and the \textit{w/o Names} descriptions, we obtained better comprehension results for the \textit{Flow} descriptions generally. However, best privacy inference results were obtained when the privacy implications were described explicitly.  Overall, these results revealed the complexity of people's data disclosure decision-making, and the importance of implication communication to help people understand DP and LDP. }

\revisionend
\section{General Discussion}
\label{sec:discussion}

The present study reports four experiments that were motivated by communicating differential privacy to facilitate users' data disclosure decisions.  In Experiments 1 and 2, we proposed different ways of describing differential privacy techniques and evaluated the effect\revision{s} of those descriptions\sout{in} \revision{on} participants' data sharing decisions (\textbf{RQ1}-\textbf{RQ2}).
In Experiments 3 and 4, reasons behind participants' data sharing decisions (\textbf{RQ3}), as well as their subjective (\textbf{RQ4}) and objective (\textbf{RQ5}) comprehensions of DP and LDP were examined, respectively.

\subsection{Summary of Main Results}
\mypara{Difficult to Understand the Data Perturbation Processes.} When we presented DP and LDP techniques based on definition, participants increased their data disclosure of high-sensitive information.  Participants’ data disclosure rates were\sout{better} \revision{larger} for DP than for LDP, despite \revision{the} latter providing better privacy guarantees. Moreover, participants reduced their data \sout{participation disclosure} \revision{sharing} when the ``local'' aspect of LDP was emphasized.  Many participants explained that they made the sharing decision because of the described techniques, but \revision{their answers} \sout{results} of objective comprehension questions indicated that they had difficulty in understanding the privacy and utility implications and might not differentiate the benefit of differential privacy from the promise of any \revision{other} privacy technology.

\mypara{Effects of Descriptions Affording Implication Inferences.}
When privacy implication\revision{s} (i.e., whether the privacy protection relies on the trustworthiness of the company)\sout{was} \revision{were} presented, participants opted out less
and shared more with LDP relative to DP. \sout{Also, the best}\revision{Together with the highest} correct answer rate\sout{for} \revision{of} privacy inference\revision{s obtained with implication descriptions, it}\sout{of attackers} indicate\revision{s} that\sout{such} \revision{participants'} data disclosure decisions were\sout{based on} \revision{closely related to their}\sout{participants’} correct understanding of\sout{the} privacy protection.
When privacy and utility inferences were embedded within data flow descriptions, participants increased their correct answer rates for objective comprehension questions, indicating that descriptions affording implication inferences facilitate participants’ comprehension of differential privacy.

\mypara{Primary Concern for Information Sensitivity.} Our results also revealed that information sensitivity is an important moderator for people’s data disclosure decisions.  On average, participants’ data disclosure rates of the high-sensitive questions\sout{was} \revision{were} 20\% less than the low-sensitive questions in the first two experiments. Participants, especially those who have medical conditions, revealed that they worried about \revision{the} negative consequences of data leakage or misuse of medical\revision{-}related information, thus chose not to share.

\subsection{Data Disclosure Decision-Making}
The effect of information sensitivity on data sharing decisions implies distribution differences between the collected low-sensitive and high-sensitive information. Such differences are informative to differential privacy algorithm\revision{s} or deployment\revision{s} in which such effect has not been accounted.

Besides information sensitivity, we also examined the framing effect to understand people's data disclosure decisions.  With a loss framing presented in Pilot Study 1, participants opted out less for \revision{the} low-sensitive questions. When a\sout{n extra} gain framing was presented in Experiments 1 and 2, participants only increased their data disclosure for \revision{the} low-sensitive questions. Thus, our results indicate that the bounded rationality of privacy decisions~\cite{acquisti2005privacy,kahneman1979prospect} was qualified by the sensitivity of information, providing an explanation for the\sout{framing effect} differences obtained in prior studies~\cite{adjerid2013sleights,gluck2016short}.

Extra factors impacting data sharing decisions were revealed in \revision{the} qualitative results of Experiment 3, which we grouped into two categories: context-dependent and trustworthiness-related.  When making data disclosure decisions, participants took \textit{personal contexts} into consideration, e.g., whether they have medical conditions, have been hacked before, or heard about reports of user data breach.  Some participants' decisions also factored in \revision{the} \textit{trustworthiness of the technique or the company}, e.g., whether they believe differential privacy \sout{are} \revision{is} tested enough or the company's intention to collect users' data.

Thus, privacy\revision{-}related decisions are multi-facet~\cite{Solove2008}.  To\revision{wards} effectively communicat\sout{e}\revision{ing} differential privacy\sout{in order} to facilitate \revision{users'} data disclosure, both general factors\sout{,} and users' specific concerns should be understood and addressed.

\subsection{Differential Privacy Communication}
Our results revealed implication descriptions as one effective way to communicate differential privacy to laypeople.  However, compared to the privacy aspect, participants still had difficulty in understanding the utility cost\revision{s}\sout{for} \revision{of} DP and LDP.  We conjecture that this is mainly because the privacy implications correspond to\sout{participants'} \revision{people's} privacy concerns\sout{,} \revision{(e.g., data breach from attackers).}\sout{,}\sout{meeting} \revision{Thus, the implication descriptions meet} their expectation of privacy protection techniques~\cite{lin2012expectation}.  Since people consider utility when making data sharing decisions, future work should examine other formats, e.g., \revision{graphs,}\sout{visual presentation} which can intuitively illustrate accuracy loss in helping people understand utility cost.  Also, to understand possible\sout{trust} \revision{trustworthiness} gap between DP and LDP, future work could include evaluation results or third\revision{-}party reports \revision{about DP and LDP}.

\sout{Participants' data disclosure decisions were impacted by specific wording, e.g., ``local'', w} With definition descriptions but not implication descriptions, \revision{participants' data disclosure decisions were impacted by specific wording, e.g., ``local'',} suggesting it may be as a result of comprehension.  That participants who did not understand privacy implications were susceptible to extraneous factors (e.g., company names) and\sout{used them} \revision{considered those factors}\sout{for} \revision{when making} data sharing decisions.

\subsection{Limitations}
We note that proper caution should be taken to generalize our fin\revision{d}ing to other settings.  First, we examined the effect of descriptions in data sharing decisions mainly based on the stronger privacy promise of LDP than DP.  Other factors, such as utility cost, might render DP and LDP not strict alternatives during preference decisions. \sout{But}\revision{However,} \sout{utility cost}comprehension \revision{of utility cost} showed no \revision{significant} difference between DP and LDP across \textit{w/o Names}, \textit{Imp.}, and \textit{Flow} descriptions.
Thus, any impact of utility cost should have \revision{a} similar effect on the obtained results.

Second, instead of answering the survey questions directly, participants indicated their willingness of data sharing in a hypothetical setting, limiting the ecological validity of \revision{the} current experimental design.  Note that we decided to use this role-play method to protect participants' privacy.  Prior studies showed that people's stated intentions and actual behavior sometimes differ~\cite{norberg2007privacy}.\sout{But} \revision{Yet,} we replicated the well-known framing effects using the health app setting and the hypothetical questions.  Thus, we are confident about our findings. Also, we were mainly interested in \revision{the} comparison between conditions,\sout{the} \revision{so any} effect of \revision{the} role-play can cancel out.

Third, we recruited MTurk workers who tended to be younger, better educated, and put more value on information privacy~\cite{kang2014privacy}.  Thus, our results may represent population having more privacy concerns than the broader U.S. public.

\section{Conclusions}
\label{sec:conc}

Differential privacy techniques are currently being transitioned from academic to industry.  Across different approaches of textual descriptions, our study shows that descriptions affording privacy and utility implications can facilitate people's data sharing decisions and their comprehension of DP and LDP techniques.  Thus, our work highlights the importance of user-centered deployment of differential privacy but also sheds light on \revision{the} challenges\sout{from the} \revision{for} usability\sout{aspect} \revision{research and studies}.
{
\bibliographystyle{abbrv}
\bibliography{bibs/abbrev0,bibs/Ninghui,bibs/privacy,bibs/password}
}

\appendices

\section{Survey Instruments}
\label{sec:all_ins}
\sout{In this appendix, we list the textual descriptions and questions asked in experiments Group 1. The materials are formatted differently than what appeared in the survey seen by the participants. The three-step role play and differential privacy description were also used in Experiments 3 and 4.}

\revisionstart
\subsection{\revision{Experiments 1, 2A \& 2B}}
\sout{We present the survey details (italicized parts) using the \textit{LDP w/o Names} condition in Experiment 1 as an example. Some procedures were described in the square brackets.}

\textit{In the current information age, everyone faces one question: Will you share your personal information in return for a product, service, or other benefits?} \revision{\textcolor{gray}{\sout{[This sentence was not presented in the \textit{Control} condition]}[the gain framing]}}

\textit{The purpose of this study is to understand what kind of information you are willing to share with a health app, and how you would like your data to be used.
}
\textit{For this survey, suppose:}
\textit{1) you just download a health app (Orange Health) and start to use it};
\textit{2) to ensure appropriate health suggestions and recommendations, the app asks you to provide some information, for example, your age and gender for accurate recommendation of daily calorie intake};
\textit{3) the app server also requests permission to access and collect information to provide you better user experience, for example, the information you shared will be used to train some machine learning algorithms at the server, which will then be used to provide more accurate suggestions for all the users}. \revision{\textcolor{gray}{[the three-step role play]}}

\textit{To respect your personal information privacy and ensure best user experience, the data shared with the app will be collected via the local differential privacy (LDP) technique. LDP protects users’ privacy by adding random noise to each response that users give such that the probability that any user’s attribute is inferred is similar as he or she is opt-out for the data collection. LDP has been used by companies such as Google and Apple.} \revision{\textcolor{gray}{[the \textit{LDP} description condition]}}

\textit{Please select which of the following description is correct about local differential privacy (LDP).}
\textbullet{ \textit{A privacy protection technique that adds random noise to the aggregated data (e.g., average age) collected from groups of users, such that the users’ privacy can be protected in the same way as they are opt-out for the data collection.}}
\textbullet{ \textit{A privacy protection technique that adds random noise to each response that the users provided such that the users’ privacy can be protected in the same way as they are opt-out for the data collection.}}
\textbullet{ \textit{It has not been implemented by any organization or company yet.}}
\textbullet{ \textit{None of the above is correct.}}
\textbullet{ \textit{I prefer not to answer.}}
\revision{\textcolor{gray}{[The \textit{LDP} description was presented again if participants did not answer the above check question correctly.  Then participants\sout{ saw 14 survey questions presented with a smartphone layout in a randomized order.  For each question, they} decided their\sout{ answer and selected how they would like their answer being used} answers for $14$ data sharing decisions with the following question.]}}
\textit{How would you like your answer to the following question being used?
}
\textbullet{ \textit{Only used by the app locally}}
\textbullet{ \textit{Used by the app locally and the server with LDP}}
\textbullet{ \textit{Neither used by the app locally nor the server with LDP}}
\textbullet{ \textit{I prefer not to answer}}
\revision{\textcolor{gray}{[After the demographic questions, participants evaluated their trust\sout{.  Besides LDP or DP, trust of the health app and the app server were evaluated in each condition with the following question,} of all following items: A \{the health app, the app server, LDP/DP\}.]}}
\textit{Please indicate your agreement with the following description:
	I trust\sout{LDP} \revision{\{A\}} to protect my personal information privacy.
}
\textbullet{ \textit{Strongly Disagree}}
\textbullet{ \textit{Mostly Disagree}}
\textbullet{ \textit{Somewhat Disagree}}
\textbullet{ \textit{Neither Agree Nor Disagree}}
\textbullet{ \textit{Somewhat Agree}}
\textbullet{ \textit{Mostly Agree}}
\textbullet{ \textit{Strongly Agree}}

\subsection{Experiment 3}
\textit{The purpose of this study is to evaluate your willingness to share information given one privacy protection technique and to understand why you decide so.}
\textcolor{gray}{[The three-step role play was presented.]}
\textit{To respect your personal information privacy and ensure better user experience, the data shared with the app will be collected via a privacy protection technique. }
\textit{Next we will present you a description of the privacy protection technique. Please read it carefully.}
\textcolor{gray}{[After viewing the \textit{LDP} description, participants answered the question below.]} \textit{Given the privacy protection provided by the technique, will you share your personal information (e.g., date of birth, family medical record, income level, substance use, surgery record, diagnostic record, current medication) with the app server?}
\textbullet{ \textit{Yes}}
\textbullet{ \textit{No}}
\textbullet{ \textit{I prefer not to answer}}
\textcolor{gray}{[Based on participants' answer, they then replied one open question.]
}
\textit{Please briefly explain why you did not want/would like to share your personal data given the described privacy protection technique.}
\textcolor{gray}{[easy-to-comprehend rating]}
\textit{Please indicate your agreement with the following description on a 7-point Likert Scale:
	(1: Strongly disagree; 7: Strongly agree).
	The prior description of the privacy protection technique was easy for me to understand. }
\textbullet{ \textit{Strongly Disagree}}
\textbullet{ \textit{Mostly Disagree}}
\textbullet{ \textit{Somewhat Disagree}}
\textbullet{ \textit{Neither Agree Nor Disagree}}
\textbullet{ \textit{Somewhat Agree}}
\textbullet{ \textit{Mostly Agree}}
\textbullet{ \textit{Strongly Agree}} \textcolor{gray}{[For participants who gave a rating smaller than 4, we asked them the following question]}
\textit{You indicated that the description of DP was not easy to understand. Please highlight the words or sentences that are difficult for you to understand.}

\subsection{Experiment 4}

\textit{The purpose of this study is to evaluate your understanding of a privacy protection technique based on the given description.}
\textcolor{gray}{[The three-step role play was presented.]}
\textit{To respect your personal information privacy and ensure better user experience, the data shared with the app will be collected via a privacy protection technique. }
\textit{Next we will present the description of the privacy protection technique. Please read it carefully and then answer several questions.}
\textcolor{gray}{[Then participants viewed one description, answered the five questions listed in Appendix C, and gave the easy-to-comprehend rating.]}

\revisionend

\section{Pilot Study 1}
\label{sec:exp1}
We validated the health app data collection setting as testbed by examining individuals' data disclosure decision of survey questions as a function of question sensitivity (\textit{high}, \textit{low}) and framing effect (\textit{Control}, \textit{Loss Framing}).  The question sensitivity was varied within subjects, and the framing effect was varied between subjects.
We predicted that participants' data disclosure for the high-sensitive questions would be less than that for the low-sensitive questions, and participants in the \textit{Loss Framing} condition would have more privacy concerns, thus would reduce data disclosure compared to the \textit{Control}.

\subsection{Participants and Stimuli}

$219$ Amazon MTurk workers completed the online survey.  The procedure for both conditions was the same as that in\sout{Section~\ref{sec:exp_overview}} \revision{Experiment 1} except that \sout{a} description about loss due to privacy concern was presented at the beginning of the \textit{Loss Framing} condition\sout{, before the scenario description}.
The extra description\sout{for} \revision{of} the \textit{Loss Framing} condition is as follows:

\begin{itemize}[leftmargin=*]
	\item
	      \underline{\textit{Loss Framing}}: \textit{\textit{In the current information age, everyone faces one question: Will you protect your personal information in sacrifice\sout{for} \revision{of} a product, service, or other benefits?}}
\end{itemize}

\subsection{Results}
Participants were excluded from data analysis using the same criteria as main experiments (see Table~\ref{tbl:exc}).
Results of $103$ participants from the \textit{Control} and $101$ from the \textit{Loss Framing} were included in the data analyses.

45.6\% of the participants were female, and their ages ranged from $18$ to over $50$ years, with 87.7\% between 18 and 44 years. 73.5\% were college students or professionals who had associates, bachelors, or higher degrees. 75.9\% of the participants claimed that they do not have a degree or work experience in computer science or related fields. The demographic distributions were similar between conditions and showed \revision{a} similar pattern as the main experiments.

\begin{table}[ht]
	\caption{Each option selection results for high-sensitive and low-sensitive questions in each condition of Pilot study 1. }
	\label{tbl:result_pilot1}
	\resizebox{0.48\textwidth}{!}{
		\begin{tabular}{lcccc}
			\hline
			\textbf{Condition}            & \textbf{Question Sensitivity} & \textbf{Opt Out} & \textbf{Local Only} & \textbf{Both} \\ \hline
			\multirow{2}{*}{Control}      & low-sensitive                 & 5.1\%            & 32.2\%              & 62.7\%        \\ \cline{2-5}
			                              & high-sensitive                & 16.1\%           & 45.5\%              & 38.4\%        \\\hline
			\multirow{2}{*}{Loss Framing} & low-sensitive                 & 2.8\%            & 37.8\%              & 59.4\%        \\ \cline{2-5}
			                              & high-sensitive                & 17.0\%           & 51.3\%              & 31.7\%        \\ \hline
		\end{tabular}
	}
\end{table}

\textit{Opt out} decision, selection for \textit{Local only} option, and choice of \textit{Both} option collapsed across participants (see Table~\ref{tbl:result_pilot1}) were entered into a $2$ (question sensitivity: \textit{low-sensitive}, \textit{high-sensitive}) $\times$ 2 (condition: \textit{Control}, \textit{Loss Framing}) chi-squared tests with a significance level of .05, respectively.  Post-hoc tests with Bonferroni corrections~\cite{bland1995multiple} were performed, testing all pairwise comparisons with corrected \textit{p}-values for possible inflation.  We mainly report the statistics for the significant effects.  Please refer to Table~\ref{tbl:stat_1} for all statistical tests results.

\textbf{Opt out rate.} Participants opted out more for the high-sensitive \revision{questions} (16.5\%) than for the low-sensitive questions (4.0\%),  $\chi_{(1)}^2= 120.5, \textit{p} < .001$.  The two-way interaction of sensitivity $\times$ condition was significant, $\chi_{(1)}^2= 4.66, \textit{p} = .031$. Between two conditions, participants' opt-out rates were similar for the high-sensitive questions (\textit{Control}: 16.1\%, \textit{Loss Framing}: 16.9\%).\sout{But} \revision{However,} for the low-sensitive questions, participants in the \textit{Loss Framing} condition (2.8\%) opted out less than those in the \textit{Control} condition (5.1\%), $\textit{$p_{adj}$} = .031$.  The results suggest that the loss framing made participants think more of ``product, service, or other benefits'' but limited to the low-sensitive information.

\textbf{Local only selection rate.} Participants selected more \textit{Local only} option for the high-sensitive \revision{questions} (48.4\%) than for the low-sensitive questions (34.9\%), $\chi_{(1)}^2= 52.55, \textit{p} < .001$.  The selection\sout{ratio} \revision{rates} for the \textit{Loss Framing} condition (44.6\%) was higher than that for the \textit{Control} condition (38.8\%), $\chi_{(1)}^2= 9.37, \textit{p} = .002$. \sout{But} \revision{Nevertheless,} the two-way interaction of question sensitivity $\times$ condition was not significant.  Thus, participants generally preferred\sout{the} high-sensitive information to be used by the app locally, and such preference was relatively independent from the framing effect.

\textbf{Both selection rate.} The results were in agreement with those of \textit{Local only} selection.  Participants chose more \textit{Both} option for the low-sensitive \revision{questions} (61.0\%) than for the high-sensitive questions (35.1\%), $\chi_{(1)}^2 = 192.02, \textit{p} < .001$.  Relative to the \textit{Control} (50.6\%), participants in the \textit{Loss Framing} selected less \textit{Both} option (45.5\%), $\chi_{(1)}^2 = 6.97, \textit{p} = .008$.  But the main effect of question sensitivity did not interact with condition.

\textbf{Trust evaluation.} Trust evaluation collapsed across participants were entered into 2 (\sout{device}\revision{data use}: \textit{the local app}, \textit{the app server}) $\times$ 2 (condition: \textit{Control}, \textit{Loss Framing}) chi-squared tests.
Only the main effect of condition was significant, $\chi_{(1)}^2 = 4.71, \textit{p} = .029$.  Participants in the \textit{Control}  showed more trust (62.1\%) than those in the \textit{Loss Framing} (50.9\%).

\subsection{Discussion}

In \sout{the} Pilot Study 1, participants showed less willingness to share\sout{the} high-sensitive \revision{information} than\sout{the} low-sensitive\sout{questions} \revision{information}.  When the loss framing was presented, participants' data disclosure was reduced regardless of question\revision{s'} sensitivity. Participants opted out less for the low-sensitive questions with the loss framing, indicating a risk seeking qualified in terms of question sensitivity.  Thus, we obtained the effect of question sensitivity and the\sout{frame} \revision{framing} effect~\cite{adjerid2013sleights,bilogrevic2016if,gluck2016short}, confirming the health app data collection setting and hypothetical willingness to disclose personal information as one testbed to evaluate privacy decisions.

\section{Pilot Study 2}
\label{sec:exp4a}
To make sure the two new descriptions and five comprehension questions are understandable to the participants, we conducted a pilot study with $20$ participants on Amazon MTurk.  Using a between-subject design, half of the participants were randomly assigned into the \textit{DP Flow} condition and the other half into the \textit{LDP Flow} condition.

\subsection{Participants, Stimuli, and Procedure} At the beginning of the study, we made it clear to the participants that we were interested in \revision{(}1) their understanding of a privacy protection technique based on the given description, and \revision{(}2)\sout{to} whether the description is clear and the survey questions are understandable.  Participants were randomly assigned to one of the descriptions.  After introducing the three steps of role play similar\sout{as} \revision{to} prior experiments, we presented the \textit{DP Flow} or \revision{the} \textit{LDP Flow} description.
Following the description, participants in each condition answered five questions evaluating their comprehension of DP\sout{and} \revision{or} LDP implications from the perspectives of privacy and accuracy. The five questions were:

\begin{itemize}[leftmargin=*]
	\item
	      \underline{\textit{Q1:}}
	      \textit{Suppose that you have answered truthfully the questions presented by the app, and the answers were collected using the privacy protection technique explained earlier.  If an attacker gets access to the database of the health app company, will the attacker be able to see your real answer?}
	\item
	      \underline{\textit{Q2:}}
	      \textit{Suppose that you have answered truthfully the questions presented by the app, and the answers were collected using the privacy protection technique explained earlier.   For employees within the health app company, will they be able to see your real answer?}
	\item
	      \underline{\textit{Q3:}}
	      \textit{Suppose that you have answered truthfully the questions presented by the app, and the answers were collected using the privacy protection technique explained earlier.  For the third party companies with which the health app company shared data, will they be able to see the real answer that you submitted?}
	\item
	      \underline{\textit{Q4:}}
	      \textit{With the modification from the privacy protection technique, the accuracy of summary results obtained by the health app company will become \underline{\hspace{1cm}} if compared to results without the privacy protection technique (\textbf{compared to the true results [without the privacy protection technique]}).}
	\item
	      \underline{\textit{Q5:}}
	      \textit{Suppose that you shared your information, such as your family medical history, with the health app. With the modification from the privacy protection technique, will the results still be useful for the third-party companies with which the health app company share data? }

\end{itemize}
The options are ``Yes'', ``No'', ``Unsure'' , or ``Prefer not to answer'' except Q4, whose options are ``Better'', ``Worse (correct answer for DP \& LDP)'', ``No change'',  ``Unsure'', or ``Prefer not to answer''.
After each question, participants also rated whether the question description was easy to comprehend on\sout{a} \revision{the} 7-point Likert Scale.  For participants who gave ratings smaller than 4, they were then asked to describe which part or parts of the description are hard to understand and briefly explain the reasons.  The five questions were presented randomly.   We also randomized the options except ``Unsure'' and ``Prefer not to answer'' for each question.  After the five questions, we also asked participants to indicate their agreement on whether the description of the privacy protection technique was easy to comprehend on the 7-point Likert Scale.  For rating\revision{s} lower than 4, we asked participants to describe which part or parts of the questions are hard to understand and briefly explain the reasons.  In the end, participants answered questions about their demographics.

\subsection{Results}
55\% of the participants were female.  15\% of them were less than 24 years old. 70\% of them were the ages of 25 to 44.  The rest 15\% were between the age of 45 to 54.  65\% of them have college or higher degrees.  35\% of the participants had \revision{a} high school degree.   65\% of them indicated that they did not have \revision{a} computer science background.

\begin{table}[ht]
	\caption{Correct answer rate and easy-to-comprehend rating for each inference question\revision{, as well as}\sout{and} average rating\sout{s} of each description in Pilot Study 2.}
	\label{tbl:result_pilot2}
	\resizebox{0.48\textwidth}{!}{
		\begin{tabular}{lcccc}
			\hline
			\multirow{2}{*}{\textbf{Question}} & \multicolumn{2}{c}{\textbf{Correct Rate}} & \multicolumn{2}{c}{\begin{tabular}[c]{@{}c@{}}\textbf{Easy-to-Comprehend}\\  \textbf{Rating}\end{tabular}}                                           \\ \cline{2-5}
			                                   & \textbf{DP Flow}.                         & \textbf{LDP Flow}                               & \textbf{DP Flow } & \textbf{LDP Flow  } \\ \hline
			Privacy\_Attacker                  & 50.0\%                                    & 80.0\%                                          & 6.2               & 5.8                 \\ \hline
			Privacy\_Employee                  & 60.0\%                                    & 70.0\%                                          & 5.6               & 6.4                 \\ \hline
			Privacy\_Third Party               & 50.0\%                                    & 90.0\%                                          & 5.7               & 6.1                 \\ \hline
			Utility\_Cost                      & 0.0\%                                     & 10.0\%                                          & 5                 & 5.2                 \\ \hline
			Utility\_Third Party               & 90.0\%                                    & 80.0\%                                          & 6.1               & 6.1                 \\ \hline
			Description                        & NA                                        & NA                                              & 6.2               & 6.1                 \\ \hline
		\end{tabular}
	}
\end{table}

Correct answer rate of each question for both descriptions are shown in Table~\ref{tbl:result_pilot2}.  Generally\revision{,} participants could understand and answered correctly better than \revision{the} chance for all questions except for the question of utility cost.   Among $20$ participants, four of them thought Q4 (Utility Cost) was hard to understand, e.g., ``\textit{The initial description of the privacy protection technique doesn't mention anything about comparing results with the original data, so I don't really know the answer to the question}'', and ``\textit{I'm not sure if there is really enough information to know whether the answers will be better or not.  The explanation is very vague.}''
Participants believed both descriptions and all five questions were easy to understand.  \revision{Based on the obtained results,} \sout{W}we modified the descriptions (see details in Table~\ref{tbl:tech}) and Q4 (see the bold part)\sout{based on the obtained results}.

\section{Additional Results}
\label{sec:add_result}
We provide\sout{participants} exclusion summary \revision{of participants} for all\sout{quantitative} experiments in Table~\ref{tbl:exc}; DP and LDP descriptions proposed in the present study and descriptions from companies are shown in Tables~\ref{tbl:tech}; statistical test results \revision{of each Pilot study and experiment} in Tables~\ref{tbl:stat_pilot1},~\ref{tbl:stat_1},~\ref{tbl:stat_2a},~\ref{tbl:stat_2b}, and~\ref{tbl:stat_4}, respectively.

\begin{table}[ht]
	\begin{center}
		\caption{\small{Statistical test results of Pilot Study 1.}}
		\label{tbl:stat_pilot1}
		\resizebox{0.48\textwidth}{!}{\begin{tabular}{|r|c|c|c|c|c|c|}
				\hline
				\multicolumn{1}{|c|}{\multirow{2}{*}{{Term}}}            & \multicolumn{2}{c|}{{Local Only}}         & \multicolumn{2}{c|}{{Both}}               & \multicolumn{2}{c|}{{Opt Out}}                                                                                            \\ \cline{2-7}
				                                                         & \textit{{$\chi^2$}}                       & \textit{{p}}                              & \textit{{$\chi^2$}}                           & \textit{{p}}             & \textit{{$\chi^2$}} & \textit{{p}}             \\ \hline
				\multicolumn{1}{|l|}{{Question Sensitivity}}             & \textbf{52.55}                            & \textbf{\textless{}.001}                  & \textbf{192.02}                               & \textbf{\textless{}.001} & \textbf{120.5}      & \textbf{\textless{}.001} \\ \cline{1-7}
				\multicolumn{1}{|l|}{{Condition}}                        & \textbf{9.37}                             & \textbf{.002}                             & \textbf{6.97}                                 & \textbf{.008}            & \textless{}1.0      &                          \\ \cline{1-7}
				\multicolumn{1}{|l|}{{Question Sensitivity * Condition}} & \textless{}1.0                            &                                           & 1.04                                          & 0.309                    & \textbf{4.66}       & \textbf{.031}            \\ \cline{1-7}
				\multicolumn{1}{|l|}{Low-Sensitive vs. High-Sensitive}   & \multicolumn{6}{c|}{}                                                                                                                                                                                             \\ \cline{1-7}
				\multicolumn{1}{|r|}{Control (Con.)}                     & \multicolumn{2}{c|}{\multirow{2}{*}{N/A}} & \multicolumn{2}{c|}{\multirow{2}{*}{N/A}} & \multicolumn{2}{c|}{\textbf{\textless{}.001}}                                                                             \\ \cline{1-1} \cline{6-7}
				\multicolumn{1}{|r|}{Loss Framing}                       & \multicolumn{2}{c|}{}                     & \multicolumn{2}{c|}{}                     & \multicolumn{2}{c|}{\textbf{\textless{}.001}}                                                                             \\ \cline{1-7}
				\multicolumn{1}{|l|}{Low-Sensitive}                      & \multicolumn{6}{c|}{}                                                                                                                                                                                             \\ \cline{1-7}
				\multicolumn{1}{|r|}{Con. vs. Loss Framing}              & \multicolumn{2}{c|}{N/A}                  & \multicolumn{2}{c|}{N/A}                  & \multicolumn{2}{c|}{\textbf{.031}}                                                                                        \\ \cline{1-7}
				\multicolumn{1}{|l|}{High-Sensitive}                     & \multicolumn{6}{c|}{}                                                                                                                                                                                             \\ \cline{1-7}
				\multicolumn{1}{|r|}{Con. vs. Loss Framing}              & \multicolumn{2}{c|}{N/A}                  & \multicolumn{2}{c|}{N/A}                  & \multicolumn{2}{c|}{\textless{}1.0}                                                                                       \\ \hline
			\end{tabular}}
	\end{center}
\end{table}

\begin{table}[ht]
	\begin{center}
		\caption{\small{Statistical test results of Experiment 1.}}
		\label{tbl:stat_1}
		\resizebox{0.48\textwidth}{!}{\begin{tabular}{|r|c|c|c|c|c|c|}
				\hline
				\multicolumn{1}{|c|}{\multirow{2}{*}{{Term}}}            & \multicolumn{2}{c|}{{Local Only}}             & \multicolumn{2}{c|}{{Both}}                   & \multicolumn{2}{c|}{{Opt Out}}                                                                                        \\ \cline{2-7}
				                                                         & \textit{{$\chi^2$}}                           & \textit{{p}}                                  & \textit{{$\chi^2$}}                       & \textit{{p}}             & \textit{{$\chi^2$}} & \textit{{p}}             \\ \hline
				\multicolumn{1}{|l|}{{Question Sensitivity}}             & \textbf{43.38}                                & \textbf{\textless{}.001}                      & \textbf{325.65}                           & \textbf{\textless{}.001} & \textbf{290.0}      & \textbf{\textless{}.001} \\ \cline{1-7}
				\multicolumn{1}{|l|}{{Condition}}                        & \textbf{40.31}                                & \textbf{\textless{}.001}                      & \textbf{30.31}                            & \textbf{\textless{}.001} & 4.95                & .175                     \\ \cline{1-7}
				Con vs. Gain Framing                                     & \multicolumn{2}{c|}{\textbf{\textless{}.001}} & \multicolumn{2}{c|}{\textbf{.012}}            & \multicolumn{2}{c|}{\multirow{6}{*}{N/A}}                                                                             \\ \cline{1-5}
				Con vs. DP                                               & \multicolumn{2}{c|}{\textbf{\textless{}.001}} & \multicolumn{2}{c|}{\textbf{\textless{}.001}} & \multicolumn{2}{c|}{}                                                                                                 \\ \cline{1-5}
				Con vs. LDP                                              & \multicolumn{2}{c|}{.101}                     & \multicolumn{2}{c|}{\textbf{.017}}            & \multicolumn{2}{c|}{}                                                                                                 \\ \cline{1-5}
				Gain Framing vs. DP                                      & \multicolumn{2}{c|}{.380}                     & \multicolumn{2}{c|}{.165}                     & \multicolumn{2}{c|}{}                                                                                                 \\ \cline{1-5}
				Gain Framing vs. LDP                                     & \multicolumn{2}{c|}{.499}                     & \multicolumn{2}{c|}{\textgreater{}.999}       & \multicolumn{2}{c|}{}                                                                                                 \\ \cline{1-5}
				DP vs. LDP                                               & \multicolumn{2}{c|}{\textbf{\textless{}.001}} & \multicolumn{2}{c|}{\textbf{.045}}            & \multicolumn{2}{c|}{}                                                                                                 \\ \cline{1-7}
				\multicolumn{1}{|l|}{{Question Sensitivity * Condition}} & \textbf{21.59}                                & \textbf{\textless{}.001}                      & \textbf{22.45}                            & \textbf{\textless{}.001} & 3.03                & .388                     \\ \cline{1-7}
				\multicolumn{1}{|l|}{Low-Sensitive vs. High-Sensitive}   & \multicolumn{6}{l|}{}                                                                                                                                                                                                 \\ \cline{1-7}
				Control                                                  & \multicolumn{2}{c|}{\textbf{\textless{}.001}} & \multicolumn{2}{c|}{\textbf{\textless{}.001}} & \multicolumn{2}{c|}{\multirow{4}{*}{N/A}}                                                                             \\ \cline{1-5}
				Gain Framing                                             & \multicolumn{2}{c|}{\textbf{\textless{}.001}} & \multicolumn{2}{c|}{\textbf{\textless{}.001}} & \multicolumn{2}{c|}{}                                                                                                 \\ \cline{1-5}
				DP                                                       & \multicolumn{2}{c|}{.113}                     & \multicolumn{2}{c|}{\textbf{\textless{}.001}} & \multicolumn{2}{c|}{}                                                                                                 \\ \cline{1-5}
				LDP                                                      & \multicolumn{2}{c|}{.113}                     & \multicolumn{2}{c|}{\textbf{\textless{}.001}} & \multicolumn{2}{c|}{}                                                                                                 \\ \cline{1-7}
				\multicolumn{1}{|l|}{Low-Sensitive}                      & \multicolumn{6}{l|}{}                                                                                                                                                                                                 \\ \cline{1-7}
				Con vs. Gain Framing                                     & \multicolumn{2}{c|}{\textbf{\textless{}.001}} & \multicolumn{2}{c|}{\textbf{\textless{}.001}} & \multicolumn{2}{c|}{\multirow{6}{*}{N/A}}                                                                             \\ \cline{1-5}
				Con vs. DP                                               & \multicolumn{2}{c|}{\textbf{.025}}            & \multicolumn{2}{c|}{\textbf{.018}}            & \multicolumn{2}{c|}{}                                                                                                 \\ \cline{1-5}
				Con vs. LDP                                              & \multicolumn{2}{c|}{\textgreater{}.999}       & \multicolumn{2}{c|}{\textgreater{}.999}       & \multicolumn{2}{c|}{}                                                                                                 \\ \cline{1-5}
				Gain Framing vs. DP                                      & \multicolumn{2}{c|}{.283}                     & \multicolumn{2}{c|}{.843}                     & \multicolumn{2}{c|}{}                                                                                                 \\ \cline{1-5}
				Gain Framing vs. LDP                                     & \multicolumn{2}{c|}{\textbf{.001}}            & \multicolumn{2}{c|}{\textbf{.013}}            & \multicolumn{2}{c|}{}                                                                                                 \\ \cline{1-5}
				DP vs. LDP                                               & \multicolumn{2}{c|}{\textbf{.047}}            & \multicolumn{2}{c|}{.409}                     & \multicolumn{2}{c|}{}                                                                                                 \\ \cline{1-7}
				\multicolumn{1}{|l|}{High-Sensitive}                     & \multicolumn{6}{l|}{}                                                                                                                                                                                                 \\ \cline{1-7}
				Con vs. Gain Framing                                     & \multicolumn{2}{c|}{\textgreater{}.999}       & \multicolumn{2}{c|}{\textgreater{}.999}       & \multicolumn{2}{c|}{\multirow{6}{*}{N/A}}                                                                             \\ \cline{1-5}
				Con vs. DP                                               & \multicolumn{2}{c|}{\textbf{\textless{}.001}} & \multicolumn{2}{c|}{\textbf{\textless{}.001}} & \multicolumn{2}{c|}{}                                                                                                 \\ \cline{1-5}
				Con vs. LDP                                              & \multicolumn{2}{c|}{\textbf{.021}}            & \multicolumn{2}{c|}{\textbf{.017}}            & \multicolumn{2}{c|}{}                                                                                                 \\ \cline{1-5}
				Gain Framing vs. DP                                      & \multicolumn{2}{c|}{\textbf{\textless{}.001}} & \multicolumn{2}{c|}{\textbf{\textless{}.001}} & \multicolumn{2}{c|}{}                                                                                                 \\ \cline{1-5}
				Gain Framing vs. LDP                                     & \multicolumn{2}{c|}{.557}                     & \multicolumn{2}{c|}{\textbf{.048}}            & \multicolumn{2}{c|}{}                                                                                                 \\ \cline{1-5}
				DP vs. LDP                                               & \multicolumn{2}{c|}{\textbf{.024}}            & \multicolumn{2}{c|}{.258}                     & \multicolumn{2}{c|}{}                                                                                                 \\ \hline
			\end{tabular}}
	\end{center}
\end{table}

\begin{table}[ht]
	\begin{center}
		\caption{\small{Statistical test results of Experiment 2A.}}
		\label{tbl:stat_2a}
		\resizebox{0.5\textwidth}{!}{\begin{tabular}{|r|c|c|c|c|c|c|}
				\hline
				\multicolumn{1}{|c|}{\multirow{2}{*}{{Term}}}          & \multicolumn{2}{c|}{{Local Only}}             & \multicolumn{2}{c|}{{Both}}                   & \multicolumn{2}{c|}{{Opt Out}}                                                                                            \\ \cline{2-7}
				                                                       & \textit{{$\chi^2$}}                           & \textit{{p}}                                  & \textit{{$\chi^2$}}                           & \textit{{p}}             & \textit{{$\chi^2$}} & \textit{{p}}             \\ \hline
				\multicolumn{1}{|l|}{Question Sensitivity}             & \textbf{41.54}                                & \textbf{\textless{}.001}                      & \textbf{331.56}                               & \textbf{\textless{}.001} & \textbf{372.68}     & \textbf{\textless{}.001} \\ \cline{1-7}
				\multicolumn{1}{|l|}{Condition}                        & 3.88                                          & .274                                          & \textless{}1.0                                &                          & 3.91                & .271                     \\ \cline{1-7}
				LDP w/o Names vs. DP w/o Names                         & \multicolumn{2}{c|}{\multirow{6}{*}{N/A}}     & \multicolumn{2}{c|}{\multirow{6}{*}{N/A}}     & \multicolumn{2}{c|}{\multirow{6}{*}{N/A}}                                                                                 \\ \cline{1-1}
				LDP w/o Names vs. LDP Comp. w/o Names                  & \multicolumn{2}{c|}{}                         & \multicolumn{2}{c|}{}                         & \multicolumn{2}{c|}{}                                                                                                     \\ \cline{1-1}
				LDP w/o Names vs. LDP Comp.                            & \multicolumn{2}{c|}{}                         & \multicolumn{2}{c|}{}                         & \multicolumn{2}{c|}{}                                                                                                     \\ \cline{1-1}
				LDP Comp. vs. LDP Comp. w/o Names                      & \multicolumn{2}{c|}{}                         & \multicolumn{2}{c|}{}                         & \multicolumn{2}{c|}{}                                                                                                     \\ \cline{1-1}
				DP w/o Names vs. LDP Comp. w/o Names                   & \multicolumn{2}{c|}{}                         & \multicolumn{2}{c|}{}                         & \multicolumn{2}{c|}{}                                                                                                     \\ \cline{1-1}
				DP w/o names vs. LDP Comp.                             & \multicolumn{2}{c|}{}                         & \multicolumn{2}{c|}{}                         & \multicolumn{2}{c|}{}                                                                                                     \\ \cline{1-7}
				\multicolumn{1}{|l|}{Question Sensitivity * Condition} & \textbf{12.74}                                & \textbf{.005}                                 & \textbf{9.06}                                 & \textbf{.029}            & \textbf{7.88}       & \textbf{.048}            \\ \cline{1-7}
				\multicolumn{1}{|l|}{Low-Sensitive vs. High-Sensitive} & \multicolumn{6}{l|}{}                                                                                                                                                                                                     \\ \cline{1-7}
				DP w/o Names                                           & \multicolumn{2}{c|}{.452}                     & \multicolumn{2}{c|}{\textbf{\textless{}.001}} & \multicolumn{2}{c|}{\textbf{\textless{}.001}}                                                                             \\ \cline{1-7}
				LDP w/o Names                                          & \multicolumn{2}{c|}{\textbf{.019}}            & \multicolumn{2}{c|}{\textbf{\textless{}.001}} & \multicolumn{2}{c|}{\textbf{\textless{}.001}}                                                                             \\ \cline{1-7}
				LDP Comp. w/o Names                                    & \multicolumn{2}{c|}{\textbf{\textless{}.001}} & \multicolumn{2}{c|}{\textbf{\textless{}.001}} & \multicolumn{2}{c|}{\textbf{\textless{}.001}}                                                                             \\ \cline{1-7}
				LDP Comp.                                              & \multicolumn{2}{c|}{\textbf{.004}}            & \multicolumn{2}{c|}{\textbf{\textless{}.001}} & \multicolumn{2}{c|}{\textbf{\textless{}.001}}                                                                             \\ \cline{1-7}
				\multicolumn{1}{|l|}{Low-Sensitive}                    & \multicolumn{6}{l|}{}                                                                                                                                                                                                     \\ \cline{1-7}
				LDP w/o Names vs. DP w/o Names                         & \multicolumn{2}{c|}{.850}                     & \multicolumn{2}{c|}{.270}                     & \multicolumn{2}{c|}{.892}                                                                                                 \\ \cline{1-7}
				LDP w/o Names vs. LDP Comp. w/o Names                  & \multicolumn{2}{c|}{\textgreater{}.999}       & \multicolumn{2}{c|}{\textgreater{}.999}       & \multicolumn{2}{c|}{\textgreater{}.999}                                                                                   \\ \cline{1-7}
				LDP w/o Names vs. LDP Comp.                            & \multicolumn{2}{c|}{\textgreater{}.999}       & \multicolumn{2}{c|}{\textgreater{}.999}       & \multicolumn{2}{c|}{.660}                                                                                                 \\ \cline{1-7}
				LDP Comp. vs. LDP Comp. w/o Names                      & \multicolumn{2}{c|}{\textgreater{}.999}       & \multicolumn{2}{c|}{\textgreater{}.999}       & \multicolumn{2}{c|}{\textgreater{}.999}                                                                                   \\ \cline{1-7}
				DP w/o Names vs. LDP Comp. w/o Names                   & \multicolumn{2}{c|}{.356}                     & \multicolumn{2}{c|}{\textgreater{}.999}       & \multicolumn{2}{c|}{\textgreater{}.999}                                                                                   \\ \cline{1-7}
				DP w/o names vs. LDP Comp.                             & \multicolumn{2}{c|}{\textgreater{}.999}       & \multicolumn{2}{c|}{.366}                     & \multicolumn{2}{c|}{\textgreater{}.999}                                                                                   \\ \cline{1-7}
				\multicolumn{1}{|l|}{High-Sensitive}                   & \multicolumn{6}{l|}{}                                                                                                                                                                                                     \\ \cline{1-7}
				LDP w/o Names vs. DP w/o Names                         & \multicolumn{2}{c|}{\textgreater{}.999}       & \multicolumn{2}{c|}{\textgreater{}.999}       & \multicolumn{2}{c|}{.681}                                                                                                 \\ \cline{1-7}
				LDP w/o Names vs. LDP Comp. w/o Names                  & \multicolumn{2}{c|}{\textbf{.014}}            & \multicolumn{2}{c|}{\textgreater{}.999}       & \multicolumn{2}{c|}{\textbf{.020}}                                                                                        \\ \cline{1-7}
				LDP w/o Names vs. LDP Comp.                            & \multicolumn{2}{c|}{\textgreater{}.999}       & \multicolumn{2}{c|}{\textgreater{}.999}       & \multicolumn{2}{c|}{.544}                                                                                                 \\ \cline{1-7}
				LDP Comp. vs. LDP Comp. w/o Names                      & \multicolumn{2}{c|}{.208}                     & \multicolumn{2}{c|}{\textgreater{}.999}       & \multicolumn{2}{c|}{\textgreater{}.999}                                                                                   \\ \cline{1-7}
				DP w/o Names vs. LDP Comp. w/o Names                   & \multicolumn{2}{c|}{\textbf{.017}}            & \multicolumn{2}{c|}{\textgreater{}.999}       & \multicolumn{2}{c|}{\textgreater{}.999}                                                                                   \\ \cline{1-7}
				DP w/o names vs. LDP Comp.                             & \multicolumn{2}{c|}{\textgreater{}.999}       & \multicolumn{2}{c|}{.248}                     & \multicolumn{2}{c|}{\textgreater{}.999}                                                                                   \\ \hline
			\end{tabular}}
	\end{center}
\end{table}

\begin{table}[ht]
	\begin{center}
		\caption{\small{Statistical test results of Experiment 2B.}}
		\label{tbl:stat_2b}
		\resizebox{0.48\textwidth}{!}{\begin{tabular}{|r|c|c|c|c|c|c|}
				\hline
				\multicolumn{1}{|c|}{\multirow{2}{*}{{Term}}}          & \multicolumn{2}{c|}{{Local Only}}         & \multicolumn{2}{c|}{{Both}}                   & \multicolumn{2}{c|}{{Opt Out}}                                                                                        \\ \cline{2-7}
				                                                       & \textit{{$\chi^2$}}                       & \textit{{p}}                                  & \textit{{$\chi^2$}}                       & \textit{{p}}             & \textit{{$\chi^2$}} & \textit{{p}}             \\ \hline
				\multicolumn{1}{|l|}{Question Sensitivity}             & \textbf{22.58}                            & \textbf{\textless{}.001}                      & \textbf{206.02}                           & \textbf{\textless{}.001} & \textbf{229.49}     & \textbf{\textless{}.001} \\ \cline{1-7}
				\multicolumn{1}{|l|}{Condition}                        & \textbf{12.36}                            & \textbf{.002}                                 & \textbf{29.2}                             & \textbf{\textless{}.001} & \textbf{10.77}      & \textbf{.005}            \\ \cline{1-7}
				LDP Imp. vs. DP Imp.                                   & \multicolumn{2}{c|}{\textbf{.049}}        & \multicolumn{2}{c|}{\textbf{\textless{}.001}} & \multicolumn{2}{c|}{\textbf{.026}}                                                                                    \\ \cline{1-7}
				LDP Imp. vs. LDP Imp. w/o Local                        & \multicolumn{2}{c|}{\textgreater{}.999}   & \multicolumn{2}{c|}{\textgreater{}.999}       & \multicolumn{2}{c|}{\textgreater{}.999}                                                                               \\ \cline{1-7}
				DP Imp. vs. LDP Imp. w/o Local                         & \multicolumn{2}{c|}{\textbf{.002}}        & \multicolumn{2}{c|}{\textbf{\textless{}.001}} & \multicolumn{2}{c|}{\textbf{.012}}                                                                                    \\ \cline{1-7}
				Question Sensitivity * Condition                       & \textless{}1.0                            &                                               & 1.36                                      & .508                     & 3.61                & .164                     \\ \cline{1-7}
				\multicolumn{1}{|l|}{Low-Sensitive vs. High-Sensitive} & \multicolumn{6}{l|}{}                                                                                                                                                                                             \\ \cline{1-7}
				DP Imp.                                                & \multicolumn{2}{c|}{\multirow{3}{*}{N/A}} & \multicolumn{2}{c|}{\multirow{3}{*}{N/A}}     & \multicolumn{2}{c|}{\multirow{3}{*}{N/A}}                                                                             \\ \cline{1-1}
				LDP Imp.                                               & \multicolumn{2}{c|}{}                     & \multicolumn{2}{c|}{}                         & \multicolumn{2}{c|}{}                                                                                                 \\ \cline{1-1}
				LDP Imp. w/o Local                                     & \multicolumn{2}{c|}{}                     & \multicolumn{2}{c|}{}                         & \multicolumn{2}{c|}{}                                                                                                 \\ \cline{1-7}
				\multicolumn{1}{|l|}{Low-Sensitive}                    & \multicolumn{6}{l|}{}                                                                                                                                                                                             \\ \cline{1-7}
				LDP Imp. vs. DP Imp.                                   & \multicolumn{2}{c|}{\multirow{3}{*}{N/A}} & \multicolumn{2}{c|}{\multirow{3}{*}{N/A}}     & \multicolumn{2}{c|}{\multirow{3}{*}{N/A}}                                                                             \\ \cline{1-1} LDP Imp. vs. LDP Imp. w/o Local & \multicolumn{2}{c|}{} & \multicolumn{2}{c|}{} & \multicolumn{2}{c|}{} \\ \cline{1-1} DP Imp.vs. LDP Imp. w/o Local & \multicolumn{2}{c|}{} & \multicolumn{2}{c|}{} & \multicolumn{2}{c|}{} \\ \cline{1-7}
				\multicolumn{1}{|l|}{High-Sensitive}                   & \multicolumn{6}{l|}{}                                                                                                                                                                                             \\ \cline{1-7}
				LDP Imp. vs. DP Imp.                                   & \multicolumn{2}{c|}{\multirow{3}{*}{N/A}} & \multicolumn{2}{c|}{\multirow{3}{*}{N/A}}     & \multicolumn{2}{c|}{\multirow{3}{*}{N/A}}                                                                             \\ \cline{1-1} LDP Imp. vs. LDP Imp. w/o Local & \multicolumn{2}{c|}{} & \multicolumn{2}{c|}{} & \multicolumn{2}{c|}{} \\ \cline{1-1} DP Imp. vs. LDP Imp. w/o Local & \multicolumn{2}{c|}{} & \multicolumn{2}{c|}{} & \multicolumn{2}{c|}{} \\ \hline
			\end{tabular}}
	\end{center}
\end{table}

\begin{table}[ht]
	\begin{center}
		\caption{\small{Statistical test results of Experiment 4.}}
		\label{tbl:stat_4}
		\resizebox{0.48\textwidth}{!}{\begin{tabular}{|l|l|c|c|}
				\hline
				{Question}                                 & {Term}                                   & {$\chi^2$}                                    & {\textit{p}}                                  \\ \hline
				\multirow{10}{*}{Q1. Privacy\_Attackers}   & Technique                                & \multicolumn{1}{c|}{\textbf{9.49}}            & \multicolumn{1}{c|}{\textbf{.002}}            \\ \cline{2-4}
				                                           & Description                              & \multicolumn{1}{c|}{\textbf{102.3}}           & \multicolumn{1}{c|}{\textless{}\textbf{.001}} \\ \cline{2-4}
				                                           & \multicolumn{1}{r|}{w/o Names  vs. Imp.} & \multicolumn{2}{c|}{\textless{}\textbf{.001}}                                                 \\ \cline{2-4}
				                                           & \multicolumn{1}{r|}{w/o Names  vs. Flow} & \multicolumn{2}{c|}{\textless{}\textbf{.001}}                                                 \\ \cline{2-4}
				                                           & \multicolumn{1}{r|}{Imp. Vs. Flow}       & \multicolumn{2}{c|}{\textbf{.002}}                                                            \\ \cline{2-4}
				                                           & Technique * Description                  & \textbf{15.91}                                & \textless{}\textbf{.001}                      \\ \cline{2-4}
				                                           & DP vs. LDP                               & \multicolumn{2}{l|}{}                                                                         \\ \cline{2-4}
				                                           & \multicolumn{1}{r|}{w/o Names}           & \multicolumn{2}{c|}{\textbf{.004}}                                                            \\ \cline{2-4}
				                                           & \multicolumn{1}{r|}{Imp.}                & \multicolumn{2}{c|}{.107}                                                                     \\ \cline{2-4}
				                                           & \multicolumn{1}{r|}{Flow}                & \multicolumn{2}{c|}{\textbf{.001}}                                                            \\ \cline{1-4}
				\multirow{10}{*}{Q2. Privacy\_Employees}   & Technique                                & \textbf{4.09}                                 & \textbf{.043}                                 \\ \cline{2-4}
				                                           & Description                              & \textbf{45.6}                                 & \textless{}\textbf{.001}                      \\ \cline{2-4}
				                                           & \multicolumn{1}{r|}{w/o Names  vs. Imp.} & \multicolumn{2}{c|}{\textbf{.042}}                                                            \\ \cline{2-4}
				                                           & \multicolumn{1}{r|}{w/o Names vs. Flow}  & \multicolumn{2}{c|}{\textless{}\textbf{.001}}                                                 \\ \cline{2-4}
				                                           & \multicolumn{1}{r|}{Imp. vs. Flow}       & \multicolumn{2}{c|}{\textbf{.001}}                                                            \\ \cline{2-4}
				                                           & Technique * Description                  & \textbf{10.77}                                & \textbf{.005}                                 \\ \cline{2-4}
				                                           & DP vs. LDP                               & \multicolumn{2}{l|}{}                                                                         \\ \cline{2-4}
				                                           & \multicolumn{1}{r|}{w/o Names}           & \multicolumn{2}{c|}{\textgreater{}.999}                                                       \\ \cline{2-4}
				                                           & \multicolumn{1}{r|}{Imp.}                & \multicolumn{2}{c|}{\textgreater{}.999}                                                       \\ \cline{2-4}
				                                           & \multicolumn{1}{r|}{Flow}                & \multicolumn{2}{c|}{\textbf{.001}}                                                            \\ \cline{1-4}
				\multirow{10}{*}{Q3. Privacy\_Third party} & Technique                                & \multicolumn{2}{c|}{\textless{}1.0}                                                           \\ \cline{2-4}
				                                           & Description                              & \textbf{24.77}                                & \textless{}\textbf{.001}                      \\ \cline{2-4}
				                                           & \multicolumn{1}{r|}{w/o Names  vs. Imp.} & \multicolumn{2}{c|}{.272}                                                                     \\ \cline{2-4}
				                                           & \multicolumn{1}{r|}{w/o Names  vs. Flow} & \multicolumn{2}{c|}{\textless{}\textbf{.001}}                                                 \\ \cline{2-4}
				                                           & \multicolumn{1}{r|}{Imp. Vs. Flow}       & \multicolumn{2}{c|}{\textbf{.006}}                                                            \\ \cline{2-4}
				                                           & Technique * Description                  & \textbf{12.66}                                & \textbf{.002}                                 \\ \cline{2-4}
				                                           & DP vs. LDP                               & \multicolumn{2}{l|}{}                                                                         \\ \cline{2-4}
				                                           & \multicolumn{1}{r|}{w/o Names}           & \multicolumn{2}{c|}{\textbf{.032}}                                                            \\ \cline{2-4}
				                                           & \multicolumn{1}{r|}{Imp.}                & \multicolumn{2}{c|}{.926}                                                                     \\ \cline{2-4}
				                                           & \multicolumn{1}{r|}{Flow}                & \multicolumn{2}{c|}{\textbf{.040}}                                                            \\ \cline{1-4}
				\multirow{10}{*}{Q4. Utility\_Cost}        & Technique                                & \multicolumn{2}{c|}{\textless{}1.0}                                                           \\ \cline{2-4}
				                                           & Description                              & \textbf{56.33 }                               & \textless{}\textbf{.001}                      \\ \cline{2-4}
				                                           & \multicolumn{1}{r|}{w/o Names  vs. Imp.} & \multicolumn{2}{c|}{.274}                                                                     \\ \cline{2-4}
				                                           & \multicolumn{1}{r|}{w/o Names  vs. Flow} & \multicolumn{2}{c|}{\textless{}\textbf{.001}}                                                 \\ \cline{2-4}
				                                           & \multicolumn{1}{r|}{Imp. Vs. Flow}       & \multicolumn{2}{c|}{\textless{}\textbf{.001}}                                                 \\ \cline{2-4}
				                                           & Technique * Description                  & 2.02                                          & .365                                          \\ \cline{2-4}
				                                           & DP vs. LDP                               & \multicolumn{2}{l|}{}                                                                         \\ \cline{2-4}
				                                           & \multicolumn{1}{r|}{w/o Names}           & \multicolumn{2}{c|}{NA}                                                                       \\ \cline{2-4}
				                                           & \multicolumn{1}{r|}{Imp.}                & \multicolumn{2}{c|}{NA}                                                                       \\ \cline{2-4}
				                                           & \multicolumn{1}{r|}{Flow}                & \multicolumn{2}{c|}{NA}                                                                       \\ \cline{1-4}
				\multirow{10}{*}{Q5. Utility\_Third party} & Technique                                & \textbf{4.07}                                 & \textbf{.044}                                 \\ \cline{2-4}
				                                           & Description                              & \textbf{37.32}                                & \textless{}\textbf{.001 }                     \\ \cline{2-4}
				                                           & \multicolumn{1}{r|}{w/o Names  vs. Imp.} & \multicolumn{2}{c|}{.115}                                                                     \\ \cline{2-4}
				                                           & \multicolumn{1}{r|}{w/o Names  vs. Flow} & \multicolumn{2}{c|}{\textless{}\textbf{.001}}                                                 \\ \cline{2-4}
				                                           & \multicolumn{1}{r|}{Imp. vs. Flow}       & \multicolumn{2}{c|}{\textbf{.001}}                                                            \\ \cline{2-4}
				                                           & Technique * Description                  & \textbf{8.96}                                 & \textbf{.011}                                 \\ \cline{2-4}
				                                           & DP vs. LDP                               & \multicolumn{2}{l|}{}                                                                         \\ \cline{2-4}
				                                           & \multicolumn{1}{r|}{w/o Names}           & \multicolumn{2}{c|}{.666}                                                                     \\ \cline{2-4}
				                                           & \multicolumn{1}{r|}{Imp.}                & \multicolumn{2}{c|}{\textbf{.002}}                                                            \\ \cline{2-4}
				                                           & \multicolumn{1}{r|}{Flow}                & \multicolumn{2}{c|}{.605}                                                                     \\ \hline
			\end{tabular}
		}
	\end{center}
\end{table}

\begin{table}[ht]
	\caption{\small{Summary of participants who were excluded from data analysis for all\sout{quantitative} experiments. \revision{The} number of participants before exclusion in each experiment is listed in the first column. }}
	\label{tbl:exc}
	\resizebox{0.5\textwidth}{!}{\begin{tabular}{|l|c|c|c|c|l|c|}
			\hline
			\multirow{3}{*}{\textbf{EXP.}}               & \multirow{3}{*}{\textbf{\begin{tabular}[c]{@{}c@{}}Median \\ Complete \\ Time (sec)\end{tabular}}} & \multirow{3}{*}{\textbf{\begin{tabular}[c]{@{}c@{}}Payment\\ (\$)  \end{tabular}}} & \multirow{3}{*}{\textbf{\begin{tabular}[c]{@{}c@{}}Same\\  IP\end{tabular}}} & \multirow{3}{*}{\textbf{\begin{tabular}[c]{@{}c@{}}  Time \\\textless \\ 120 sec\end{tabular}}} & \multicolumn{2}{c|}{\textbf{\begin{tabular}[c]{@{}c@{}}2nd Check Question\\ Failure\end{tabular}}}                                   \\ \cline{6-7}
			                                             &                                                       &                                                       &                                                       &                                                       & \multirow{2}{*}{\textbf{Condition}}                       & \multirow{2}{*}{\textbf{Count}} \\
			                                             &                                                       &                                                       &                                                       &                                                       &                                                           &                                 \\ \hline
			\begin{tabular}[c]{@{}l@{}}Pilot 1\\ (219)\end{tabular}                  & 264                                                   & \multirow{10}{*}{1}                                   & 4                                                     & 11                                                    & \multicolumn{2}{c|}{N/A}                                                                    \\ \cline{1-2}\cline{4-7}
			\multirow{2}{*}{\begin{tabular}[c]{@{}l@{}}EXP. 1\\ (598)\end{tabular}} & \multirow{2}{*}{289 }                                 &
			                                             & \multirow{2}{*}{12}                                   & \multirow{2}{*}{28}                                   & DP                                                    & 42                                                                                                                                                  \\ \cline{6-7}
			                                             &                                                       &                                                       &                                                       &                                                       & LDP                                                       & 51                              \\ \cline{1-2}\cline{4-7}
			\multirow{4}{*}{\begin{tabular}[c]{@{}l@{}}EXP. 2A\\ (781)\end{tabular}} & \multirow{4}{*}{328 }                                 &                                                       & \multirow{4}{*}{15}                                   & \multirow{4}{*}{19}                                   & DP w/o Names                                              & 58                              \\ \cline{6-7}
			                                             &                                                       &                                                       &                                                       &                                                       & LDP w/o Names                                             & 46                              \\ \cline{6-7}
			                                             &                                                       &                                                       &                                                       &                                                       & LDP Comp. w/o Names                                       & 30                              \\ \cline{6-7}
			                                             &                                                       &                                                       &                                                       &                                                       & LDP Comp.                                                 & 32                              \\ \cline{1-2}\cline{4-7}
			\multirow{3}{*}{\begin{tabular}[c]{@{}l@{}}EXP. 2B\\ (600)\end{tabular}} & \multirow{3}{*}{303 }                                 &                                                       & \multirow{3}{*}{9}                                    & \multirow{3}{*}{15}                                   & DP Imp.                                                   & 28                              \\ \cline{6-7}
			                                             &                                                       &                                                       &                                                       &                                                       & LDP Imp.                                                  & 43                              \\ \cline{6-7}
			                                             &                                                       &                                                       &                                                       &                                                       & LDP Imp. w/o Local                                        & 37                              \\ \hline
			\begin{tabular}[c]{@{}l@{}}EXP. 3\\ (279)\end{tabular}                  & 192                                                   & 0.75                                                  & 1                                                     & N/A                                                   & \multicolumn{2}{c|}{N/A}                                                                    \\ \hline
			\begin{tabular}[c]{@{}l@{}}Pilot 2\\ (20)\end{tabular}                  & 303                                                   & \multirow{2}{*}{1}                                    & 0                                                     & 0                                                     & \multicolumn{2}{c|}{N/A}                                                                    \\ \cline{1-2} \cline{4-7}
			\begin{tabular}[c]{@{}l@{}}EXP. 4\\ (599)\end{tabular}                  & 245                                                   &                                                       & 7                                                     & 52                                                    & \multicolumn{2}{c|}{N/A}                                                                    \\ \hline
		\end{tabular}
	}
\end{table}

\begin{table*}[ht]
	\begin{center}
		\caption{Proposed descriptions of DP and LDP techniques of each condition in each experiment. Descriptions of DP w/o Names, LDP w/o Names, and LDP Comp. w/o Names are the same as DP, LDP, and LDP Comp., respectively, except the last sentences in square brackets are deleted.  DP Flow and LDP Flow are updated based on results of Pilot Study 2 (in bold font).
		}
		\label{tbl:tech}
		\resizebox{0.8\textwidth}{!}{
			\begin{tabular}{|l|l|}
				\hline
				\textbf{Condition}                   & \textbf{Description}        \\ \hline
				\textbf{DP}                          & \begin{tabular}[c]{@{}l@{}} To respect your personal information privacy and ensure best user experience, the data shared with \\the app will be processed via the differential privacy (DP) technique.  DP protects users' privacy by \\adding random noise to aggregated data, for example, average age, such that the probability that an \\individual's information is inferred is low. [DP has been used across academia and industry, \\including Harvard University, U.S. Census Bureau, and companies such as LinkedIn and Uber.]  \end{tabular} \\ \hline
				\textbf{LDP}                         & \begin{tabular}[c]{@{}l@{}} To respect your personal information privacy and ensure best user experience, the data shared with \\the app will be collected via the local differential privacy (LDP) technique.  LDP protects users' \\privacy by adding random noise to each response that users give such that the probability that any \\user's attribute is inferred is similar as he or she is opt-out for the data collection. [LDP has been \\used by companies such as Google and Apple.]
				\end{tabular} \\ \hline
				\textbf{LDP Comp}                    & \begin{tabular}[c]{@{}l@{}} To respect your personal information privacy and ensure best user experience, the data shared with \\the app will be collected via the local differential privacy (LDP) technique. LDP protects your \\privacy by adding random noise to the raw data locally BEFORE you give the data to the company \\(raw data never leaves your device).  [LDP has been used by companies such as Google and Apple.]  \end{tabular} \\ \hline
				\textbf{DP Imp.}                     & \begin{tabular}[c]{@{}l@{}} To respect your personal information privacy and ensure best user experience, the data shared with \\the app will be processed via the differential privacy (DP) technique.  That is, the app company will \\store your data but only use the aggregated statistics with modification so that your personal \\information cannot be learned.  However, your personal information may be leaked if the company's \\database is compromised. \end{tabular} \\ \hline
				\textbf{LDP Imp.}                    & \begin{tabular}[c]{@{}l@{}}  To respect your personal information privacy and ensure best user experience, the data shared with \\the app will be processed via the local differential privacy (LDP) technique. That is, the app \\will randomly modify your data on your cellphone before sending it to the app server.  Since the \\app server stores only the modified version of your personal information, your privacy is protected \\even if the app server's database is compromised. \end{tabular} \\ \hline
				\textbf{\begin{tabular}[c]{@{}l@{}}LDP Imp.\\w/o Local\end{tabular}} & \begin{tabular}[c]{@{}l@{}}  To respect your personal information privacy and ensure best user experience, the data shared with \\the app will be processed via the differential privacy (DP) technique. That is, the app will \\randomly modify your data on your cellphone before sending it to the app server. Since the app \\server stores only the modified version of your personal information, your privacy is protected even \\if the app server's database is compromised.   \end{tabular} \\ \hline
				\textbf{DP Flow}                     & \begin{tabular}[c]{@{}l@{}} When differential privacy (DP) is used, the app sends the user's answers to the company.  These \\answers are then stored in the company's databases.  When the company wants to use the collected \\data, either internally or for sharing with third-party companies, the company sends queries to the\\ databases, applies DP techniques to modify the returned results, and uses only the modified results.  \\These modified results reveal limited information specific to each individual user.  However, by \\examining the modified answers of a large number of users, the company can still get useful \\summary results in the user population, \textbf{even though the accuracy is reduced (compared to the} \\ \textbf{case when no modification is applied)}. \end{tabular} \\ \hline
				\textbf{LDP Flow}                    & \begin{tabular}[c]{@{}l@{}}  When local differential privacy (LDP) is used, the app modifies the answers before sending them \\from the user's device to the company.  The company only sees and stores the modified version of \\each user's information, and is uncertain about each individual user's true answer.  However, by \\examining the modified answers of a large number of users, the company can still get useful \\summary results in the user population, \textbf{even though the accuracy is reduced (compared to the} \\
					\textbf{case when no modification is applied)}.\end{tabular} \\ \hline

				\hline
				\textbf{Apple}                       & \begin{tabular}[c]{@{}l@{}} Differential privacy transforms the information shared with the company before it ever leaves the \\user's device such that the company can never reproduce the true data. The differential privacy \\technology used is rooted in the idea that statistical noise that is slightly biased can mask a user’s \\individual data before it is shared with the company. If many people are submitting the same data, \\the noise that has been added can average out over large numbers of data points, and the company \\can see meaningful information emerge. \end{tabular} \\ \hline
				\textbf{Google}                      & \begin{tabular}[c]{@{}l@{}}Building on the concept of randomized response, local differential privacy (LDP) enables learning \\statistics about the behavior of users’ software while guaranteeing client privacy. LDP builds on the \\above concept, allowing software to send reports that are effectively indistinguishable from the \\results of random coin flips and are free of any unique identifiers. However, by aggregating the \\reports we can learn the common statistics that are shared by many users.  \end{tabular} \\ \hline
				\textbf{Microsoft}                   & \begin{tabular}[c]{@{}l@{}}Differential privacy is a technology that enables researchers and analysts to extract useful answers \\from databases containing personal information and, at the same time, offers strong individual \\privacy protections. This seemingly contradictory outcome is achieved by introducing relatively \\small inaccuracies in the answers provided by the system. These inaccuracies are large enough that \\they protect privacy, but small enough that the answers provided to analysts and researchers are still \\useful. \end{tabular} \\ \hline
				\textbf{Uber}                        & \begin{tabular}[c]{@{}l@{}}Differential privacy is a formal definition of privacy and is widely recognized by industry experts \\as providing strong and robust privacy assurances for individuals. In short, differential privacy allows \\general statistical analysis without revealing information about a particular individual in the data. \\Results do not even reveal whether any individual appears in the data. For this reason, differential \\privacy provides an extra layer of protection against re-identification attacks as well as attacks using \\auxiliary data.     \end{tabular} \\ \hline
				\textbf{\begin{tabular}[c]{@{}l@{}}US\\Census Bureau\end{tabular}} & \begin{tabular}[c]{@{}l@{}}Differential privacy was developed by researchers at Microsoft and is now utilized by many leading \\tech firms. There are many variants of differential privacy. The one we selected introduces controlled \\noise into the data in a manner that preserves the accuracy at higher levels of geography. Our \\differential privacy methods will be designed to preserve the utility of our legally mandated data \\products while also ensuring that every respondents’ personal information is fully protected.      \end{tabular} \\ \hline
			\end{tabular}
		}
	\end{center}
\end{table*}

\end{document}